\documentclass[aps,prd,superscriptaddress,nofootinbib,preprint]{revtex4-1}

\pdfoutput=1 

\usepackage{amsmath}
\usepackage{bm}
\usepackage{times}
\usepackage{braket}
\usepackage{color,graphicx}
\usepackage{float}
\usepackage{color}
\usepackage{hyperref}
\usepackage{multirow}

\definecolor{nicered}{rgb}{0.7,0.1,0.1}
\definecolor{nicegreen}{rgb}{0.1,0.5,0.1}
\hypersetup{colorlinks,citecolor= nicegreen,linkcolor= nicered}
\definecolor{red}{rgb}{1.0, 0, 0}

\newcommand{\beq}{\begin{equation}}
\newcommand{\eeq}{\end{equation}}
\newcommand{\bea}{\begin{eqnarray}}
\newcommand{\eea}{\end{eqnarray}}

\newcommand {\E}[1]{\times 10^{#1}}	
\newcommand {\e}[1]{\mathrm{~#1}}       
\newcommand{\mc}[1]{\mathcal{#1}}

\newcommand{\mrm}[1]{\mathrm{#1}}
\newcommand{\re}[0]{\mrm{Re}}

\begin{document}
\preprint{ZU-TH-2/15}
\preprint{DO-TH 15/04}

\title{New Physics Models Facing Lepton Flavor Violating Higgs Decays at the Percent Level}

\author{Ilja Dor\v sner} \email[Electronic address:]{dorsner@fesb.hr}
\affiliation{University of Split, Faculty of Electrical Engineering, Mechanical Engineering and Naval Architecture in Split (FESB), R.\ Bo\v skovi\' ca 32, 21 000 Split, Croatia}
\affiliation{Jo\v zef\ Stefan Institute, Jamova 39, P.\ O.\ Box 3000, 1001
  Ljubljana, Slovenia}

\author{Svjetlana Fajfer} \email[Electronic
address:]{svjetlana.fajfer@ijs.si} 
\affiliation{Department of Physics,
  University of Ljubljana, Jadranska 19, 1000 Ljubljana, Slovenia}
\affiliation{Jo\v zef\ Stefan Institute, Jamova 39, P.\ O.\ Box 3000, 1001
  Ljubljana, Slovenia}

\author{Admir Greljo} 
\email[Electronic address:]{admir@physik.uzh.ch} 
\affiliation{Physik-Institut, Universit\"at Z\"urich, CH-8057 Z\"urich, Switzerland}

\author{Jernej~F.~Kamenik} 
\email[Electronic address:]{jernej.kamenik@ijs.si} 
\affiliation{Jo\v zef\ Stefan Institute, Jamova 39, P.\ O.\ Box 3000, 1001
  Ljubljana, Slovenia}
\affiliation{Department of Physics,
  University of Ljubljana, Jadranska 19, 1000 Ljubljana, Slovenia}

\author{Nejc Ko\v snik} 
\email[Electronic address:]{nejc.kosnik@ijs.si}
\affiliation{Department of Physics,
  University of Ljubljana, Jadranska 19, 1000 Ljubljana, Slovenia}
\affiliation{Jo\v zef\ Stefan Institute, Jamova 39, P.\ O.\ Box 3000, 1001
  Ljubljana, Slovenia}
  
\author{Ivan Ni\v sand\v zi\'c}
\email[Electronic address:]{ivan.nisandzic@tu-dortmund.de}
\affiliation{Institut f\"ur Physik, Technische Universit\"at Dortmund, D-44221 Dortmund,
Germany}

\begin{abstract}
We speculate about the possible interpretations of the recently observed excess in the $h \to \tau \mu$ decay. We derive a robust lower bound on the Higgs boson coupling strength to a tau and a muon, even in presence of the most general new physics affecting other Higgs properties. Then we reevaluate complementary indirect constraints coming from low energy observables as well as from theoretical considerations. In particular, the tentative signal should lead to  $\tau \to \mu\gamma$ at rates which could be observed at Belle II. In turn we show that, barring fine-tuned cancellations, the effect can only be accommodated within models with an extended scalar sector. These general conclusions are demonstrated using a number of explicit new physics models. Finally we show how, given the $h \to \tau \mu$ signal, the current and future searches for $\mu \to e\gamma$ and $\mu \to e$ nuclear conversions unambiguously constrain the allowed rates for $h \to \tau e$.
\end{abstract}
\pacs{}
\maketitle

\section{Introduction}
The discovery of the Higgs boson at the LHC~\cite{Aad:2012tfa, Chatrchyan:2012ufa} imbues the standard model (SM) of particle physics with completeness and self-consistency. Nonetheless, its failure to account  for non-vanishing neutrino masses is one of the main motivations for considering physics beyond the SM. Incidentally, the accidental SM symmetries that prevent neutrinos from acquiring mass also completely suppress lepton flavor violating (LFV) processes. The observation of the former thus provides ample motivation for a rich experimental program to search for the latter. The CMS collaboration has recently reported a slight excess with a significance of $2.4\,\sigma$ in the search for LFV decay $h\to \tau \mu$~\cite{Khachatryan:2015kon}. The best fit for the branching ratio of the Higgs boson to $\tau \mu $ final state (summed over  $\tau^- \mu^+$ and $\tau^+ \mu^-$), assuming SM Higgs production, is found to be 
\beq
\mathcal{B}(h\to \tau \mu)=\left(0.84^{+0.39}_{-0.37}\right) \% \,.
\label{eq:Br}
\eeq
This recent hint has expectedly received significant amount of attention in the literature~\cite{Campos:2014zaa,Celis:2014roa,Sierra:2014nqa,Lee:2014rba,Heeck:2014qea,Crivellin:2015mga,deLima:2015pqa}. It is thus an imperative to either confirm or reject validity of this tantalizing hint with more data by both ATLAS and CMS experiments. At the same time, it is instructive to revisit expectations for this observable within various new physics (NP) scenarios and in particular re-evaluate the feasibility of obtaining such a large signal in light of severe indirect constraints on LFV Higgs interactions coming from low energy probes. 

\begin{figure}
\begin{centering}
\includegraphics[scale=0.65]{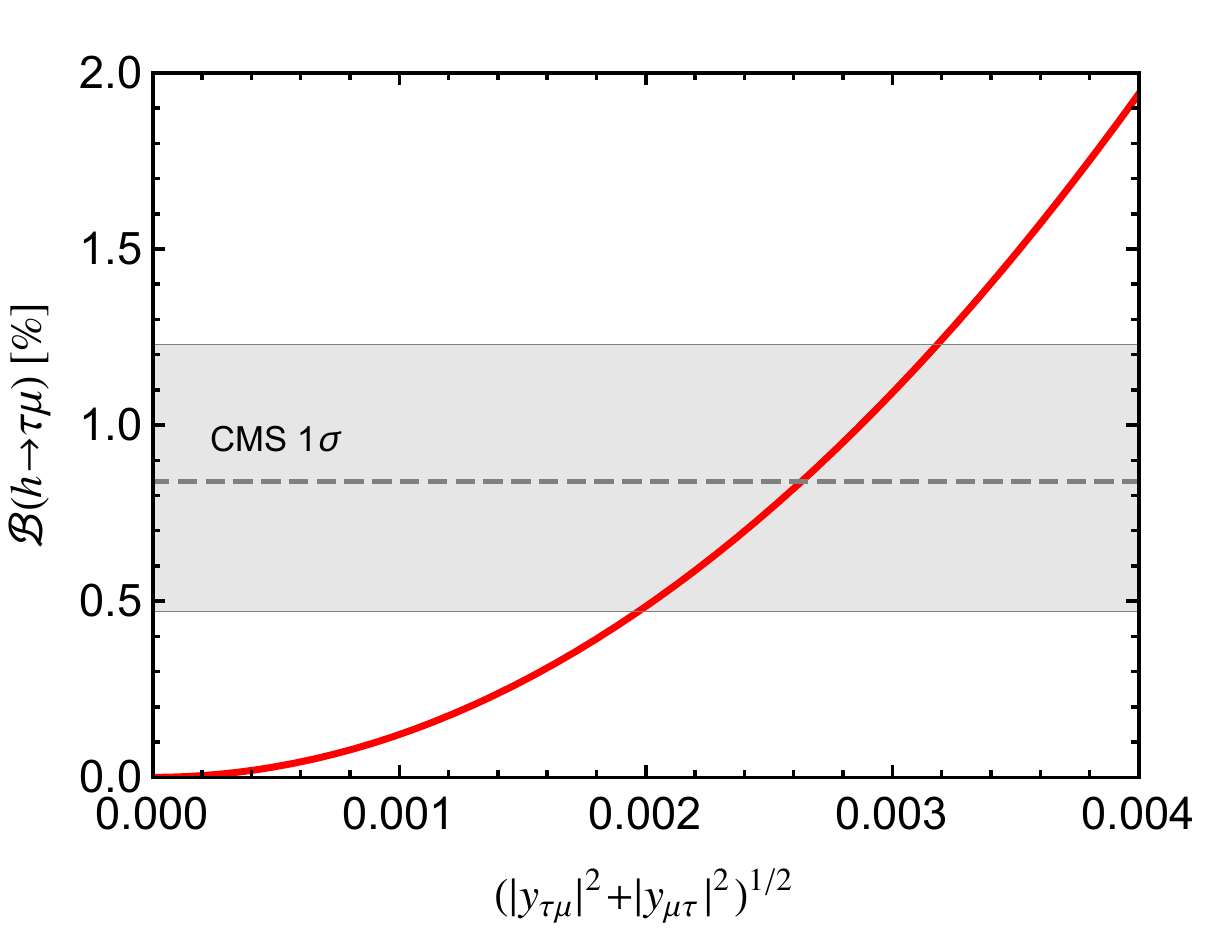} \;
\includegraphics[scale=0.64]{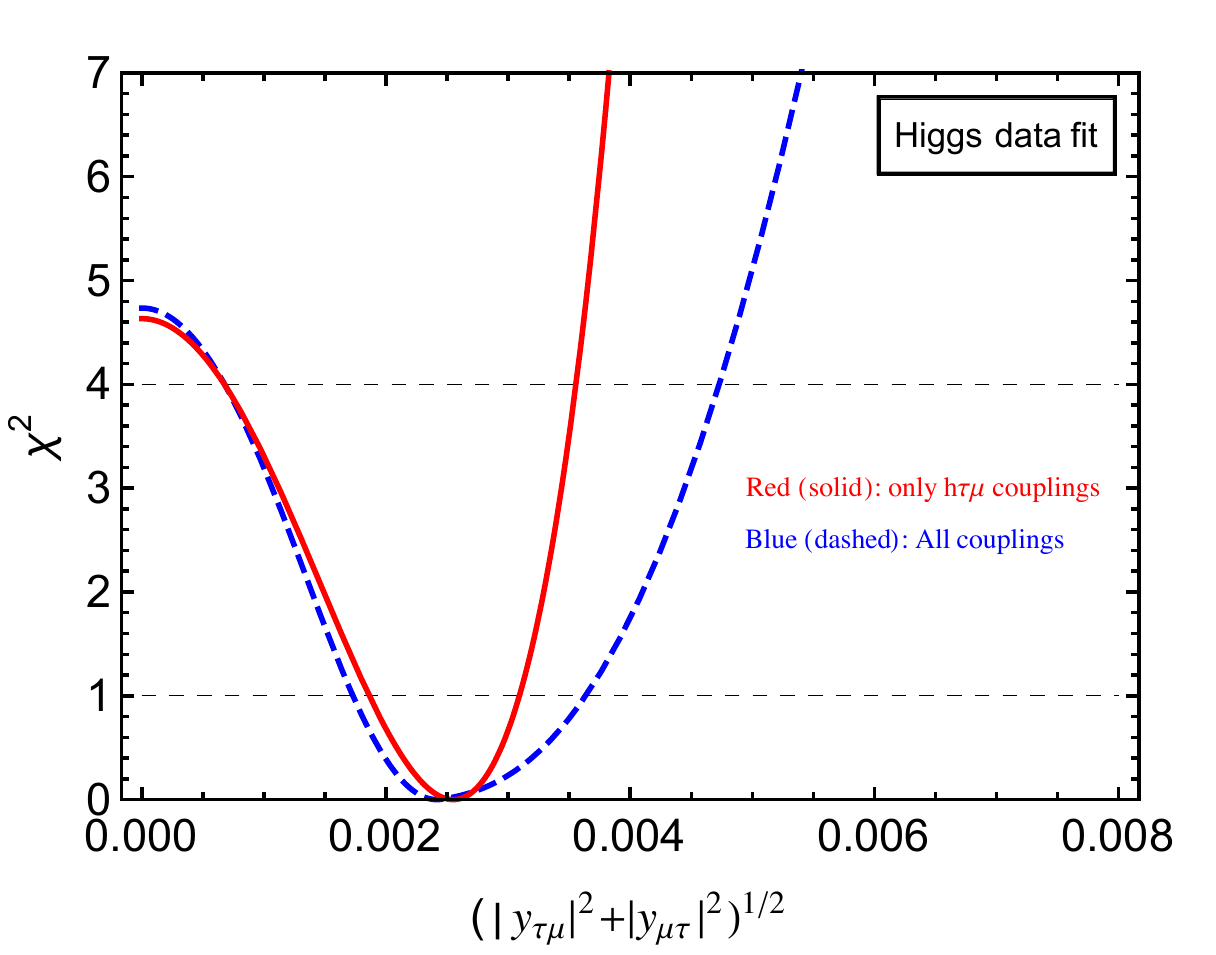}
\par\end{centering}
\caption{(Left-hand side panel) Higgs branching ratio to the $\tau\mu$ final state $\mathcal{B}(h\to \tau \mu)$ as a function of the effective $h\tau\mu$ coupling. (Right-hand side panel) The $\chi^2$ fit to the latest available LHC Higgs data including the CMS  $h\to\tau\mu$ signal: varying only the effective $h\tau \mu$ coupling (shown in red, solid curve), and varying all Higgs boson couplings (shown in blue, dashed curve).}
\label{fig:HiggsFit}
\end{figure}

Without loss of generality, one can parameterize the mass terms and Higgs boson couplings of charged leptons after electroweak symmetry breaking (EWSB) as
\beq
\mathcal L^{\rm eff.}_{Y_\ell} = - m_i \delta_{ij} \bar \ell_L^i \ell_R^j - y_{ij}\left( \bar \ell_L^i \ell_R^j\right) h + \ldots+ \rm h.c. \,,
\label{eq:LY}
\eeq
where the ellipsis denotes non-renormalizable interactions involving more than one Higgs boson and $\ell^i = e,\mu,\tau$. In the SM, the Higgs couplings are diagonal  and given by $y_{ij} = (m_i/v) \delta_{ij}$, where $v=246$\,GeV\,. On the other hand, non-zero $y_{\tau \mu}$ and/or $y_{\mu \tau}$ induce $h\to \tau \mu$ decays with a  branching ratio of
\begin{equation}
\mathcal{B}(h\to \tau \mu) = \frac{m_h}{8\pi \Gamma_{h}} \left(|y_{\tau \mu}|^2+|y_{\mu \tau}|^2\right).
\label{eq:Brhtm}
\end{equation}
Assuming that the total Higgs boson decay width ($\Gamma_{h}$) is given by its SM value enlarged only by the contribution from $h\to \tau\mu$ itself, i.e., $\Gamma_{h} = \Gamma_{h}^{\rm SM}/[1 - \mathcal{B}(h\to \tau \mu) ]$, where $\Gamma_{h}^{\rm SM} (m_h=125\,{\rm GeV}) = 4.07$\,MeV~\cite{Heinemeyer:2013tqa}, the measurement in Eq.~\eqref{eq:Br} can be interpreted as a two-sided bound on the $|y_{\tau \mu}|^2+|y_{\mu \tau}|^2$ combination of couplings. These limits read (see also the left-hand side panel of Fig.~\ref{fig:HiggsFit})
\beq
0.0019 (0.0008) < \sqrt{|y_{\tau \mu}|^2+|y_{\mu \tau}|^2} < 0.0032 (0.0036) ~ {\rm at}~ 68\%~ (95\%)~{\rm C.L.}\,.
\label{eq:htaummu1}
\eeq
In general, the experimentally measured  $h \to \tau\mu$ event yield depends not only on the values of $y_{\tau\mu}$ and $y_{\mu\tau}$, but also on other Higgs couplings contributing both to its total decay width $\Gamma_{h}$ as well as its production cross-section ($\sigma_h$). In particular, a given signal can be reproduced  for larger (smaller) values of $|y_{\tau\mu}|$ and $|y_{\mu\tau}|$ by enhancing (suppressing) $\Gamma_{h}$ and/or suppressing (enhancing) $\sigma_h$. Since both $\Gamma_{h}$ and $\sigma_h$ affect other currently measured Higgs observables, their individual effects can be disentangled by performing a global fit to all Higgs production and decay event yields at the LHC. The details of this procedure can be found in Appendix~\ref{sec:appHiggsFit}, while the resulting best fit $\chi^2$ values as functions of $\sqrt{|y_{\tau \mu}|^2+|y_{\mu \tau}|^2}$ are shown in the right-hand side panel of Fig.~\ref{fig:HiggsFit}. In particular, $\sigma_h$ is well determined by the measurements of both inclusive and separate exclusive Higgs production channels in several different decay modes. Similarly, $\Gamma_{h}$ is bounded from below by the observations (or at least indications~\cite{Chatrchyan:2013zna}) of the dominant SM Higgs decay modes ($h\to b\bar b$, $h\to WW^*$, $h\to \tau^+\tau^-$, etc.). Consequently the lower range of allowed $|y_{\tau\mu}|$ and $|y_{\mu\tau}|$ values does not change much compared to Eq.~\eqref{eq:htaummu1} when considering the global fit. On the other hand, the fact that the total Higgs decay width is currently only weakly bounded from above~\cite{Khachatryan:2014iha,ATLAS-CONF-2014-042} allows significantly larger $|y_{\tau\mu}|$ and $|y_{\mu\tau}|$ couplings to reproduce the same observed signal in the general case.\footnote{It has been argued recently~\cite{Englert:2014aca,Azatov:2014jga}, that the current total Higgs decay width measurements using off-peak data~\cite{Khachatryan:2014iha,ATLAS-CONF-2014-042}  introduce some model-specific assumptions  into the fit. We have checked that removing the total Higgs width constraint from the fit, no upper bound on $\sqrt{|y_{\tau \mu}|^2+|y_{\mu \tau}|^2}$ can be set.}  Numerically, we find
\beq
0.0017 (0.0007) < \sqrt{|y_{\tau \mu}|^2+|y_{\mu \tau}|^2} < 0.0036 (0.0047) ~ {\rm at}~ 68\%~ (95\%)~{\rm C.L.}\,.
\label{eq:htaummu1b}
\eeq

The rest of the paper is devoted to interpreting these ranges in terms of hypothetical new physics (NP) effects. In Sec.~\ref{sec:EFT} we review the model independent considerations of flavor violating Higgs couplings using an effective field theory approach. Sec.~\ref{sec:THDM} is devoted to a case study of the two Higgs doublet model (THDM) with a generic Yukawa structure (the so-called type-III THDM) that can account for the observed anomaly at tree level and be in agreement with present low energy flavor constraints. Scenario that relies on vector-like fermions to explain the data is presented in Sec.~\ref{sec:VLL}. Possibility to explain the anomaly via one-loop physics effects is discussed in Sec.~\ref{sec:LQ}, where we investigate phenomenology of scalar leptoquarks to demonstrate difficulties with this type of approach. In particular, Sec.~\ref{sec:finetune} gives a very special solution that can accommodate the anomaly through an interplay between two different one-loop effects, and satisfy constraints imposed by low energy flavor physics experiments at the price of fine-tuning. We briefly conclude in Sec.~\ref{sec:conclusions}.

\section{Model independent considerations}
\label{sec:EFT}
In the following we assume the SM contains all the relevant degrees of freedom at energies $\mathcal O({\rm few~} 100)$\,GeV.\footnote{In the minimal extension of the SM with Dirac neutrinos accounting for the observed neutrino masses, a non-vanishing $h\to \tau \mu$ rate is induced at the one-loop level. However, this contribution is negligibly small since the relevant decay amplitude is suppressed by two powers of  neutrino Yukawa couplings in addition to the tau Yukawa.} In particular, any additional heavy field can be integrated out, so that we can apply effective field theory (EFT) methods. If the electroweak symmetry is realized linearly, then current data already strongly suggest that $h$ for the most part should be a neutral CP even component of a linear combination of $Y=1/2$ hypercharge weak doublets $H_\alpha = (h_\alpha^+,v_\alpha+x_\alpha h+ \ldots)^T$, where the ellipsis denotes possible projections to other heavier scalar mass eigenstates and  $\sum_\alpha |x_\alpha|^2 \lesssim 1/2$. In general, EWSB can receive contributions from several sources (for example several Higgs multiplet condensates). However, electroweak precision tests severely constrain contributions from higher representations, so that also $\sum_\alpha v_\alpha^2 \lesssim v^2/2$.

In the decoupling limit we are considering, the contributions from the heavier Higgses are power suppressed and can be completely described in terms of higher dimensional operators. The leptonic Yukawa sector of such a theory, including the leading (dimension six) non-renormalizable operators suppressed by the EFT cut-off scale $\Lambda$, can be written as
\beq
\mathcal L_{Y_\ell} = - \lambda^\alpha_{ij} \bar L_i H_\alpha E_j - \lambda^{\prime \alpha\beta\gamma}_{ij} \frac{1}{\Lambda^2} \bar L_i H_\alpha E_j (H_{\beta}^\dagger H_\gamma) + \rm h.c.\,,
\label{eq:LYl}
\eeq
where $L_i$ and $E_i$ correspond to the leptonic weak doublets ($L_i = (\ell_L^i,\nu^i)^T$) and singlets ($E_i = \ell_R^i$), respectively, and we explicitly show the flavor indices $i,j=1,2,3$. Any additional dimension six operators coupling the Higgses to the leptons can be shown to be either redundant or not to contribute to the effective Lagrangian~\eqref{eq:LY} after EWSB~\cite{Harnik:2012pb}. We can thus identify 
\beq
y_{ij} = \frac{m_i}{v} \delta_{ij} + \epsilon_{ij}\,,
\label{eq:ye}
\eeq
where 
\beq
\epsilon =  V_L \left[\lambda^\alpha \bar v_\alpha \left( \frac{x_\alpha}{\bar v_\alpha}-1\right) + \lambda^{\prime\alpha\beta\gamma} \frac{v^2}{\Lambda^2} \bar v_\alpha \bar v_\beta \bar v_\gamma \left( \frac{x_\alpha}{\bar v_\alpha} + \frac{x_\beta}{\bar v_\beta} + \frac{x_\gamma}{\bar v_\gamma} - 1 \right) \right] V_R^\dagger\,,
\label{eq:eps}
\eeq
with $\bar v_\alpha = v_\alpha/v$ and the unitary matrices $V_{L,R}$ diagonalize the leptonic mass term as
\beq
\frac{m}{v} = V_L \left(\lambda^\alpha \bar v_\alpha + \lambda^{\prime\alpha\beta\gamma} \frac{v^2}{\Lambda^2} \bar v_\alpha  \bar v_\beta  \bar v_\gamma \right) V_R^\dagger\,. 
\label{eq:m}
\eeq
We can note immediately the two possible sources of non-vanishing $y_{\tau\mu}$ and/or $y_{\mu\tau}$.  In a theory with multiple Higgs doublets, the first term in the square brackets in Eq.~\eqref{eq:eps} can be non-vanishing if for at least two terms in the $\alpha$-sum (with non-commuting  $\lambda^\alpha$) $x_\alpha \neq \bar v_\alpha$. Thus (lepton) flavor violating Higgs boson interactions are possible even without additional NP contributions above the EFT cut-off. The simplest example of such a theory, the two Higgs doublet model (THDM) of type-III has recently been analyzed in this context in Refs.~\cite{Sierra:2014nqa,deLima:2015pqa} and we consider it in more detail in Sec.~\ref{sec:THDM}. On the other hand, in a single Higgs theory only $v_1=v/\sqrt 2$ and $x_1=1/\sqrt 2$ are non-zero so that the first term in the square brackets in Eq.~\eqref{eq:eps} vanishes identically. Thus, the observation of $h\to \tau\mu$ in this context necessarily implies the presence of the second term in Eq.~\eqref{eq:LYl}. In particular, the CMS measurement can be interpreted directly in terms of the effective NP scale of
\beq
\Lambda \simeq 4~{\rm TeV} \left[ \left( \frac{0.84\%}{\mathcal B(h\to\tau\mu)} \right) \left( |V_L \lambda^{\prime 111} V_R^\dagger|_{\tau\mu}^2+|V_L \lambda^{\prime 111} V_R^\dagger|_{\mu\tau}^2 \right) \right]^{1/4} \,.
\eeq
We observe that percent level $\mathcal B(h\to \tau\mu)$ can be accommodated within the EFT approach (containing a single light Higgs doublet) with a sizable mass gap. At the same time, new degrees of freedom generating the observed $h\to \tau\mu$ excess should be within the kinematic reach of the second LHC run.

\begin{figure}
\begin{centering}
\includegraphics[scale=1.2]{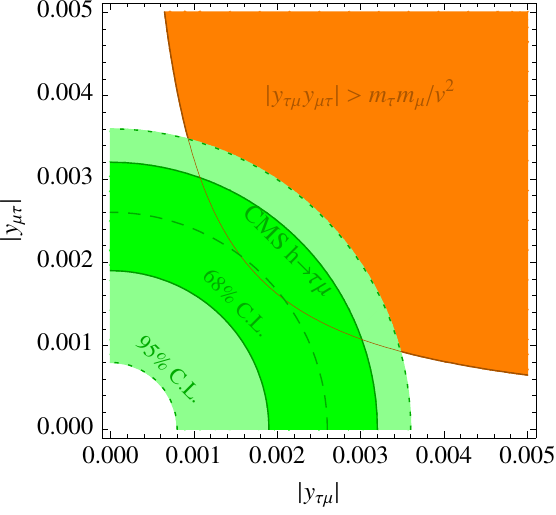} \;
\par\end{centering}
\caption{Comparison of the CMS implied range of $h\tau\mu$ couplings (green band) and the theoretical constraint imposed by the hierarchy of tau and muon masses (orange top-right region). See text for details.}
\label{fig:naturalness}
\end{figure}

Since $\lambda$ and $\lambda'$ contribute to both $\epsilon$ as well as $m$ the observed hierarchical structure of the charged lepton masses can be used to define the natural ranges of $\epsilon_{ij}$. In particular, without delicate cancellations between the various contributions in Eq.~\eqref{eq:m}, the observed hierarchy between the muon and tau lepton masses implies~\cite{Cheng:1987rs, Branco:2011iw}
\beq
\label{eq:lfv-nat}
\sqrt{|y_{\tau\mu} y_{\mu\tau}|} \lesssim \frac{\sqrt{m_\mu m_\tau}}{v} = 0.0018\,.
\eeq 
This constraint is compared to the CMS preferred range for $|y_{\tau\mu}|$ and $|y_{\mu\tau}|$ from Eq.~\eqref{eq:htaummu1} in Fig.~\ref{fig:naturalness}. We observe that the two indications are not in sharp conflict and thus the observation of $\mathcal B(h\to\tau\mu)$ at the percent level can in principle be explained by natural NP although a mild hierarchy between  $y_{\tau\mu}$ and $y_{\mu\tau}$ is preferred in this case.

\begin{figure}
\begin{centering}
\includegraphics[scale=0.5]{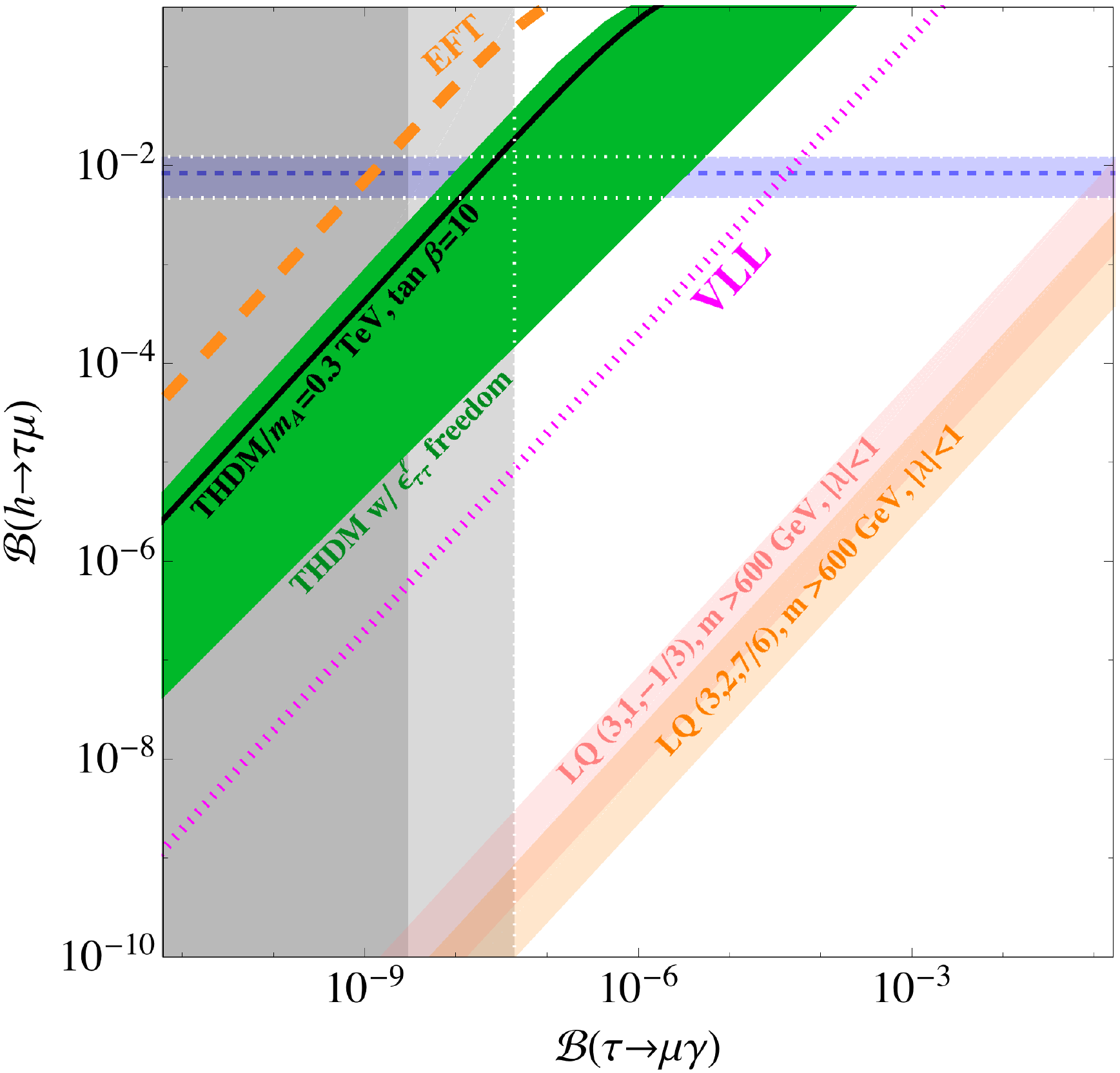} \;
\par\end{centering}
\caption{Correlation between $\mathcal B(h\to \tau\mu)$ and $\mathcal B(\tau \to \mu\gamma)$ in various NP scenarios. The present experimental result for $\mathcal B(h\to \tau\mu)$ is shown in horizontal blue band~\cite{Khachatryan:2015kon}. Current and future projections for $\mathcal B(\tau \to \mu\gamma)$ experimental sensitivity are represented with vertical light~\cite{Aubert:2009ag} and dark~\cite{Aushev:2010bq} gray bands, respectively. Superimposed are the predictions within the EFT approach (diagonal dashed orange line), in the type-III THDM (green and black bands), in models with vector-like leptons (diagonal dotted purple line) and in models with scalar leptoquarks (diagonal red and orange shaded band).  See text for details.}
\label{fig:master}
\end{figure}

Another class of model-independent constraints comes from the fact that the dimension six operators contained in $\mathcal L_{Y_\ell}$ in Eq.~\eqref{eq:LYl} will in general mix with other operators contained in the effective SM. In particular, restricting the discussion to a single Higgs setup with $H\equiv H_1$, $\lambda'$ will mix under charged lepton Yukawa renormalization into $(H^\dagger H)^3$ affecting, for example, Higgs boson pair production at the LHC. However, as shown in Refs.~\cite{Jenkins:2013zja, Jenkins:2013wua}, such mixing is suppressed by three powers of the charged lepton Yukawa matrix and is thus (baring fine-tuned cancellations) completely negligible in practice.  More phenomenologically relevant are the UV finite one- and two-loop contributions to the operators $\bar L H (\sigma \cdot B) E$ and $\bar L \tau_a H (\sigma \cdot W^a) E$, where $\sigma_{\mu\nu} = i[\gamma_\mu,\gamma_\nu]$/2, $B_{\mu\nu}$ and $W^a_{\mu\nu}$ are the hypercharge and weak isospin field strengths, respectively, and $\tau_a$ are the Pauli matrices. These operators can mediate radiative flavor violating lepton decays as well as contribute to leptonic anomalous electric and magnetic dipole moments, all of which are already tightly constrained by experiment. The most stringent constraint comes from the $\tau \to \mu\gamma$ decay~\cite{Harnik:2012pb}, mediated by the effective Lagrangian
\beq
\mathcal L_{\rm eff.} = c_L \mathcal Q_{L\gamma} + c_R \mathcal Q_{R\gamma} + \rm h.c.\,,
\label{eq:Leff}
\eeq
where $\mathcal Q_{L,R\gamma} = (e/8\pi^2) m_\tau (\bar \mu \sigma^{\alpha\beta} P_{L,R} \tau)F_{\alpha\beta}$, $P_{L,R} = (1\mp \gamma_5)/2$ and $F_{\alpha\beta}$ is the electromagnetic field strength tensor.
The Wilson coefficients $c_{L,R}$ receive comparable one- and two-loop contributions. In the experimentally justified approximation $y_{\mu\mu} \ll y_{\tau\tau}$  (both assumed real) and $m_\mu \ll m_\tau \ll m_h$, they are given by~\cite{Goudelis:2011un,Blankenburg:2012ex,Harnik:2012pb}
\begin{align}
c_L^{(\rm{1-loop})} &\simeq \frac{1}{m_h^2} y^*_{\tau\mu}  y_{\tau\tau} \left( - \frac{1}{3} + \frac{1}{4} \log \frac{m_h^2}{m_\tau^2}\right)\,, & c_R^{({\rm 1-loop})} &\simeq \frac{1}{m_h^2} y_{\mu\tau} y_{\tau\tau} \left( - \frac{1}{3} + \frac{1}{4} \log \frac{m_h^2}{m_\tau^2}\right)\,, \\
c_L^{(2-\rm loop)} &\simeq  \frac{1}{(125\,\rm GeV)^2}y^*_{\tau\mu} (0.11-0.082 y_{tt})\,,  & c_R^{(2-\rm loop)} & \simeq \frac{1}{(125\,\rm GeV)^2} y_{\mu\tau} (0.11-0.082 y_{tt})\,,
\end{align}
where $y_{tt}$ is the top quark Yukawa with the SM value of $y_{tt} = \bar m_t/v = 0.67 $ for a $\overline {\rm MS}$ top mass of $\bar m_t \simeq 164$\,GeV. The resulting EFT correlation between $\mathcal B(h\to \tau \mu)$ in Eq.~\eqref{eq:Brhtm}, and $\mathcal B(\tau \to \mu\gamma)$ given by
\beq
\mathcal B(\tau\to\mu\gamma) = \frac{\tau_\tau \alpha_{\rm EM} m_\tau^5}{64\pi^4} \left( |c_L|^2 + |c_R|^2 \right)\,,
\eeq
is shown in Fig.~\ref{fig:master} (diagonal dashed orange line), assuming SM values of all Higgs boson couplings except $y_{\tau\mu}$ and $y_{\mu\tau}$.
In the same plot, the CMS preferred range of $\mathcal B(h\to\tau\mu)$ in Eq.~\eqref{eq:Br} is displayed by the horizontal blue band, while the current ($\mathcal B(\tau\to\mu\gamma) < 4.4 \times 10^{-8}$~@~90\%~C.L.)~\cite{Aubert:2009ag} and projected future ($\mathcal B(\tau\to\mu\gamma) < 3 \times 10^{-9}$~@~90\%~C.L.)~\cite{Aushev:2010bq} indirect constraints are shaded in light and dark gray vertical bands, respectively. We observe that within the EFT approach, the CMS signal is well compatible with the non-observation of $\tau\to\mu\gamma$ at the B factories and will marginally remain so even at Belle~II. This is in contrast with the situation in most explicit NP models generating non-zero $y_{\tau\mu}$ or $y_{\mu\tau}$, as we demonstrate in Secs.~\ref{sec:THDM}--\ref{sec:LQ}\,.

Before going through explicit examples, let us consider the appearance of (lepton) flavor violating Higgs interactions from a symmetry point of view. In absence of Higgs Yukawa interactions, the SM possesses a large global flavor symmetry $\mathcal G_F = SU(3)^5$ (where we have omitted the $U(1)$ factors). In particular, the leptons transform as $L \sim (3,1)$, $E\sim (1,3)$ under $\mathcal G_\ell \equiv SU(3)_L\times SU(3)_E \in \mathcal G_F$. In the SM (without neutrino masses), the charged lepton Yukawa matrix $\lambda \sim (3,\bar 3)$ is the only source of $\mathcal G_\ell$ breaking. Consequently all lepton interactions are flavor conserving in the charged lepton mass basis.  Conversely, as also demonstrated explicitly in Eq.~\eqref{eq:eps}, the generation of lepton flavor violating Higgs interactions requires at least two non-aligned sources of lepton flavor symmetry breaking. At the tree level, there are only two possibilities: (1) one can enlarge the SM scalar sector, such that more than one Higgs doublet couples to the leptons (corresponding to the first term in Eq.~\eqref{eq:eps}); (2) one can extend the leptonic sector by vector-like fermions, whose Dirac masses and mixing terms with SM chiral fields can provide additional sources of $\mathcal G_\ell$ breaking. This leads to the appearance of the $\lambda'$ contributions after integrating out the new heavy fermionic states. Both possibilities are explored in the following sections. Example of an enlarged Higgs sector is given in Sec.~\ref{sec:THDM} whereas the vector-like fermion case is discussed in Sec.~\ref{sec:VLL}.

Of course many more possibilities exist at the loop level (e.g.~\cite{Arganda:2004bz,Arganda:2014dta}). In practice however, all such models predicting sizable LFV Higgs interactions necessarily suffer from a severe fine-tuning problem, if they are to simultaneously avoid the stringent $\tau \to\mu \gamma$ constraint. We demonstrate this on an explicit example in Sec.~\ref{sec:finetune}. A general heuristic argument however goes as follows. The operator $\bar L H E (H^\dagger H)$ has exactly the same transformation properties under both $\mathcal G_\ell$ and chiral transformations of leptons as operators $\bar L H (\sigma \cdot B) E $ and $\bar L \tau^a H (\sigma \cdot W_a) E $. Thus, the simultaneous presence of the former and absence of the latter cannot be protected by a symmetry valid in the low energy theory. If $\bar L H E (H^\dagger H)$ is generated at the loop level, it will necessarily involve charged states propagating in the loop. These charged states will couple to photons generating in turn $c_{L,R}$ in Eq.~\eqref{eq:Leff} at the same loop level as $y_{\tau\mu}$ and $y_{\mu\tau}$, respectively.  The two sets of matching amplitudes will also exhibit identical dependence on the fundamental flavor parameters of the model. The resulting expectation $c_{L,R} \sim 8\pi y_{\tau\mu,\mu\tau}/v m_\tau$ is clearly in conflict with existing experimental results, since a percent level $\mathcal B(h\to \tau\mu)$ would imply even an order of magnitude bigger $\mathcal B (\tau\to\mu\gamma)$! The accidental cancelation required in the matching procedure for the radiative decay in order to accommodate both results is thus expected to be approximately one part in $10^{3}$ at the amplitude level. As we demonstrate in Secs.~\ref{sec:THDM},~\ref{sec:VLL} and~\ref{sec:LQ}, these expectations are confirmed in explicit models.

\subsection{$h \to \tau \mu$ vs.\ $h \to \tau e$}
\label{sec:hlfv-corr-eft}
A positive experimental indication for the $h \to \tau \mu$ decay  when combined with existing stringent experimental limits on $\mu-e$ LFV processes
can be used to constrain LFV $\tau-e$ processes. In particular, the product of the $h \to \tau e$ and $h \to \tau \mu$ branching fractions is constrained from above by the rates of $\mu \to e \gamma$, and $\mu-e$ conversion on nuclei. This is due to the fact that tree level Higgs decays to $\tau \mu$ ($\tau e$) depend on $y_{\mu \tau, \tau \mu}$ ($y_{e \tau, \tau e}$) while the same sets of couplings contribute at the loop level to $\mu \to e \gamma$ and $\mu-e$ conversion via diagrams with a virtual $\tau$.

The contributions to the  $\mu \to e \gamma$ process stemming from a virtual $\tau$ are $m_\tau$ enhanced with respect to diagrams with intermediate $\mu$ or $e$ states~\cite{Harnik:2012pb} leading to
\begin{equation}
  \label{eq:muegamma-cL}
c_L^{\tau} \simeq \frac{m_\tau}{m_\mu}\, \frac{-3+4 x_\tau - x_\tau^2 - 2 \log x_\tau}{8m_h^2 (1-x_\tau)^3}\,y_{\mu \tau}^* y_{\tau e}^*\,,\qquad x_\tau = \frac{m_\tau^2}{m_h^2}\,,
\end{equation}
where we have neglected the effects of the light lepton masses while $m_\tau$ dependence is retained in its exact form. The coefficient $c_R^\tau$ is obtained from Eq.~\eqref{eq:muegamma-cL} by replacing $y_{ij} \to y_{ji}^*$. The $\mu \to e \gamma$ branching fraction is thus sensitive to a
distinct combination of the LFV Yukawas:
\begin{equation}
  \label{eq:5}
  \mc{B}(\mu \to e \gamma) \simeq \mc{B}_0^{\mu \to e \gamma} \left(|y_{\mu \tau} y_{\tau e}|^2 + |y_{\tau \mu} y_{e \tau}|^2\right)\,,\qquad \mc{B}_0^{\mu \to e \gamma} = 185\,.
\end{equation}
On the other hand, $\mu-e$ conversion on nuclei is most sensitive to vector current effective operators $(\bar e \gamma_\nu P_{L,R}\mu)\,(\bar q \gamma^\nu q)$ which probe different combinations of LFV Yukawas. The largest contribution to the vector coupling of the proton
follows from Eq.~(A17) of Ref.~\cite{Harnik:2012pb} and reads, in the limit of massless light leptons,
\begin{equation}
  \label{eq:6}
  \tilde g_{LV}^{(p)} \simeq \frac{\alpha}{2\pi}\,
\frac{-16+45 x_\tau - 36 x_\tau^2 + 7 x_\tau^3 + 6(-2+3 x_\tau)\log x_\tau }{36 m_h^2 (1-x_\tau)^4}
 \,y_{e\tau} y_{\mu \tau}^*\,.
\end{equation}
The above effective coupling is evaluated in the $q^2 \to 0$ limit, where $q$ is the momentum exchanged in the conversion process. As before the right-handed current coefficient  $g_{RV}^{(p)}$ is obtained from the above result by replacing $y_{ij}$ with $y_{ji}^*$. Note that due to strong experimental limits on $\mc{B}(\mu\to e \gamma)$  we can at present safely neglect the dipole operator $\mc{Q}_{L,R}$ contributions to $\mu-e$ conversion~\cite{Harnik:2012pb}. The relative conversion rate $\mathcal B(\mu\to e)_{\rm Au} \equiv {\Gamma(\mu\to e)_\textrm{Au}}/{\Gamma_\textrm{capture Au}}$ is then given by
\begin{equation}
  \label{eq:mueconvEFT}  
\begin{split}
\mathcal B(\mu\to e)_{\rm Au}   &\simeq \frac{{V^{(p)}}^2\,\left(|\tilde g_{LV}^{(p)}  |^2
+|\tilde g_{RV}^{(p)}  |^2\right)}{\Gamma_\textrm{capture Au}} \\
&= \mc{B}_0^{\mu e}\,\left(|y_{e\tau} y_{\mu \tau}|^2 + |y_{\tau e} y_{\tau \mu}|^2\right)\,,\qquad \mc{B}_0^{\mu e} = 4.67\E{-4}\,,
\end{split}
\end{equation}
where the coefficient $V^{(p)} = 0.0974\, m_\mu^{5/2}$ and the muon capture rate $\Gamma_\textrm{capture Au}  = 13.07\E{6}\,\e{s}^{-1}$ have been taken from Ref.~\cite{Kitano:2002mt}.

The complementary information on the LFV couplings extracted from the rates of $\mu \to e \gamma$ and $\mu-e$ conversion is sufficient to completely correlate the Higgs LFV  $h \to \tau \mu$ and $h \to \tau e$ decays:
\begin{equation}
\begin{split}
  \label{eq:taue-taumu}
  \mc{B}(h \to \tau \mu) \times   \mc{B}(h \to \tau e) &= \left[\frac{m_h}{8\pi\Gamma_h}\right]^2 \left(\frac{\mc{B}(\mu \to e \gamma)}{\mc{B}_0^{\mu \to e \gamma}} + \frac{\mathcal B(\mu\to e)_{\rm Au}}{\mc{B}_0^{\mu e}}\right) \\
&= 7.95\E{-10}\,\left[\frac{\mc{B}(\mu \to e \gamma)}{10^{-13}}\right] + 3.15\E{-4}\, \left[\frac{\mathcal B(\mu\to e)_{\rm Au}}{10^{-13}}\right]\,,
\end{split}
\end{equation}
where in the second line we have approximated $\Gamma_h \simeq \Gamma_h^{\rm SM}$.
The best experimental limit on $\mathcal B(\mu\to e)_{\rm Au} < 7\E{-13}$~(at 90\% C.L.) was achieved by the SINDRUM II Collaboration~\cite{Bertl:2006up} while the best upper bound on $\mc{B}(\mu \to e \gamma) < 5.7\E{-13}$~(at 90\% C.L.)  was recently determined by the MEG Collaboration~\cite{Adam:2013mnn}. Note that with the current experimental data the sum on the right-hand side of Eq.~\eqref{eq:taue-taumu} is completely saturated by the $\mu-e$ conversion contribution. Combining the two bounds with the central value for the $h \to \tau \mu$ branching fraction in Eq.~\eqref{eq:Br} leads to an upper bound
$  \mc{B}(h \to \tau e)  < 0.26$\,.
This is above the current indirect constraint coming from searches for $\tau \to e \gamma$~\cite{Aubert:2009ag} which reads $ {\mc{B}(h \to \tau e)  <0.19}$\,.
\begin{figure}[!htbp]
  \centering
  \includegraphics[scale=.5]{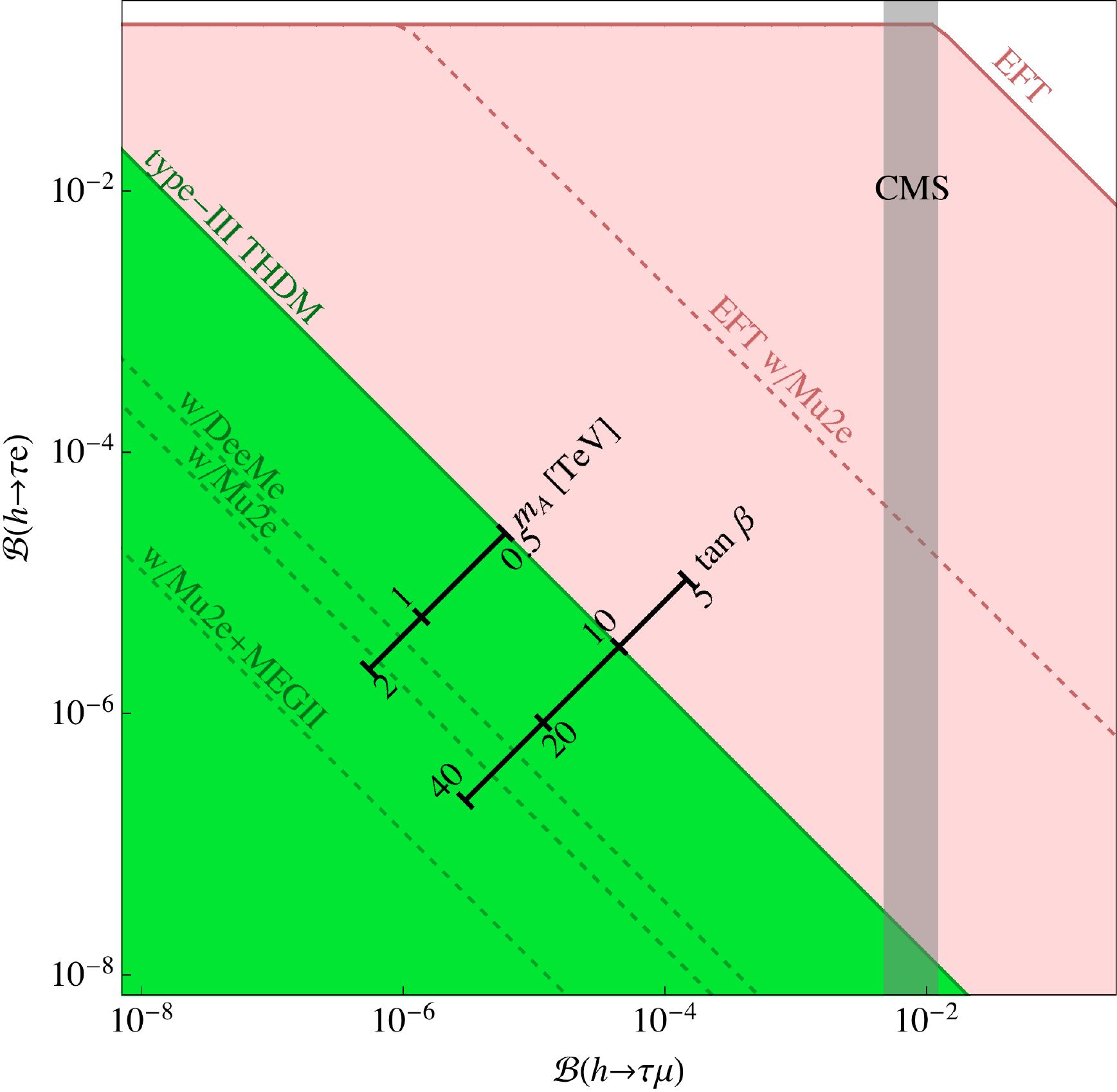}
  \caption{Allowed region in the $\mc{B}(h \to \tau e)$--$\mc{B}(h \to \tau \mu)$ plane when experimental upper bounds on $\mu \to e \gamma$ and $\mu - e$ conversion rates are taken into account. Pink region is permitted in the effective theory setting while the dashed line indicates how much the region will shrink if Mu2e and MEG II experiments see no signal events. Green region is allowed within type-III THDM
model with $m_A = 0.5\e{TeV}$ and $\tan \beta = 10$. Rulers indicate how much the region shrinks with increasing $\tan \beta$ or $m_A$, while dashed lines correspond to improved experimental upper bounds on $\mu \to e \gamma$ and $\mu-e$ as described in the text. Independently, an upper bound from $\mc{B}(\tau \to e \gamma)$ applies on the $\mc{B}(h \to \tau e)$ in the effective theory.}
  \label{fig:htaue-htaumu}
\end{figure}
In the near future the experimental sensitivity to both $\mu \to e \gamma$ and especially $\mu-e$ conversion in nuclei is expected to improve significantly. In particular the DeeMe~\cite{Aoki:2012zza} experiment aims at a sensitivity of the order of $10^{-14} $ while the Mu2e~\cite{Kutschke:2011ux} experiment could improve existing SINDRUM II bounds by four orders of magnitude. Given $\mathcal B(h\to \tau\mu)$ at the percent level, such a measurement will indirectly probe $\mathcal B(h\to \tau e )$ at the order of magnitude $10^{-5}$.

\section{Type-III Two Higgs Doublet Model}
\label{sec:THDM}
Several well motivated scenarios of physics beyond the SM (e.g. supersymmetric extensions or models addressing the strong CP problem via a PQ symmetry)
require the addition of another $SU(2)_L$ doublet scalar to the minimal SM Higgs sector (c.f.~\cite{Branco:2011iw} for a recent general review). The most extensively studied version of the
THDM is the type-II model in which
the absence of tree level flavor changing neutral currents (FCNCs)
is imposed by the requirement that only one of the doublets couples to
up-type quarks, while the other one couples to down-type quarks and charged
leptons. FCNCs in the quark sector are relegated to the one-loop level via
the interactions involving charged scalars, whose couplings to quarks
follow the CKM structure of the SM. In this model LFV is absent as in the SM due to massless neutrinos and because only a single $\lambda^\alpha$ in Eq.~\eqref{eq:LYl} is non-vanishing.

In order to accommodate LFV Higgs interactions one thus has to depart from this version and consider the THDM with a generic Yukawa structure (type-III THDM). In the following we restrict our discussion to the parameter space with a MSSM-like Higgs potential, for which the following relations hold~\cite{Crivellin:2013wna}
\begin{equation}
\begin{split}
\tan\beta &= \frac{v_{u}}{v_{d}},\quad \tan 2\alpha= \tan 2\beta\frac{m_{A}^{2}+m_{Z}^{2}}{m_{A}^{2}-m_{Z}^{2}}\,,\\
m_{H^{\pm}}^{2} &= m_{A}^{2}+m_{W}^{2}\quad m_{H}^{2} = m_{A}^{2}+m_{Z}^{2}-m_{h}^{2}\,,
\end{split}
\end{equation}
where $0<\beta<\pi/2$ is the angle of the rotation matrix that diagonalizes the mass squared matrices of charged scalars and pseudoscalars and $-\pi/2 < \alpha<0$ is an analogous angle for the neutral scalars. The vacuum expectation values of the doublets are denoted by $v_u$ and $v_d$ and $v/\sqrt{2}=\sqrt{v_u^2+v_d^2}\simeq 174$\,GeV. In our input we set the mass of the Higgs boson to the observed value $m_h\simeq 125$\,GeV and choose $\beta$ and $m_A$ to be free parameters of the scalar potential. Such a THDM can arise as a low energy effective theory of the MSSM in the limit in which all the heavy superpartners are integrated out. The LFV interactions of the Higgs boson then originate from diagrams involving heavy sleptons in the loop, c.f.~\cite{Babu:2002et, Hisano:2010es}. Note however that  generically in the MSSM, LFV radiative lepton decay contributions will be generated at the same order as discussed in Sec.~\ref{sec:EFT}, leading to tighter constraints, c.f.~\cite{Arana-Catania:2013xma}. We thus do not attempt to interpret the type-III THDM model parameters in terms of the underlying MSSM parameters, but consider them as essentially uncorrelated, only subject to experimental bounds and theoretical constraints such as perturbativity conditions.

The relevant part of the Yukawa Lagrangian in the charged lepton mass eigenstate basis is~\cite{Crivellin:2013wna}
\begin{equation}
\mathcal{L} =\frac{y_{fi}^{H_k}}{\sqrt{2}}H_k
\bar{\ell}_{L,f}\ell_{R,i}+ \frac{y_{fi}^{H^+}}{\sqrt{2}} H^+
\bar{\nu}_{L,f}\ell_{R,i} + \textrm{h.c.}\,,
\end{equation}
where the Yukawa couplings can be written as
\begin{equation}
y_{fi}^{H_{k}}=x_{d}^{k}\frac{m_{\ell_{i}}}{v_{d}}\delta_{fi}-\epsilon_{fi}^{\ell}\left(x_{d}^{k}\,\tan\beta-x_{u}^{k*}\right)\,.
\end{equation}
The terms $\epsilon^\ell_{fi}$ parametrize the off-diagonal charged lepton Yukawa couplings and are free parameters of the model. Finally, the coefficients $x_q^k$ for $H_k=(H,h,A)$ are given by~\cite{Crivellin:2013wna}
\begin{equation}
\begin{split}
x_{u}^{k} &=\left(-\sin\alpha,-\cos\alpha,i\cos\beta\right)\,, \\
x_{d}^{k} &= \left(-\cos\alpha,\sin\alpha,i\sin\beta\right)\,.
\end{split}
\end{equation}
In addition to tree level exchanges of neutral Higgses we have loop
contributions to FCNCs due to charged Higgs, whose interactions with
leptons are parameterized by the following Yukawa couplings:
\begin{equation}
  y_{fi}^{H^\pm} = \sqrt{2} \sum_{j=1}^3 \sin\beta \, V^\mathrm{PMNS}_{fj}
  \left(\frac{m_{\ell_i}}{v_d} \delta_{ji}- \epsilon^\ell_{ji} \tan \beta \right)\,.
\end{equation}

\subsection{$h \to \tau \mu$}
\begin{figure}[!th]
\begin{centering}
\includegraphics[scale=0.75]{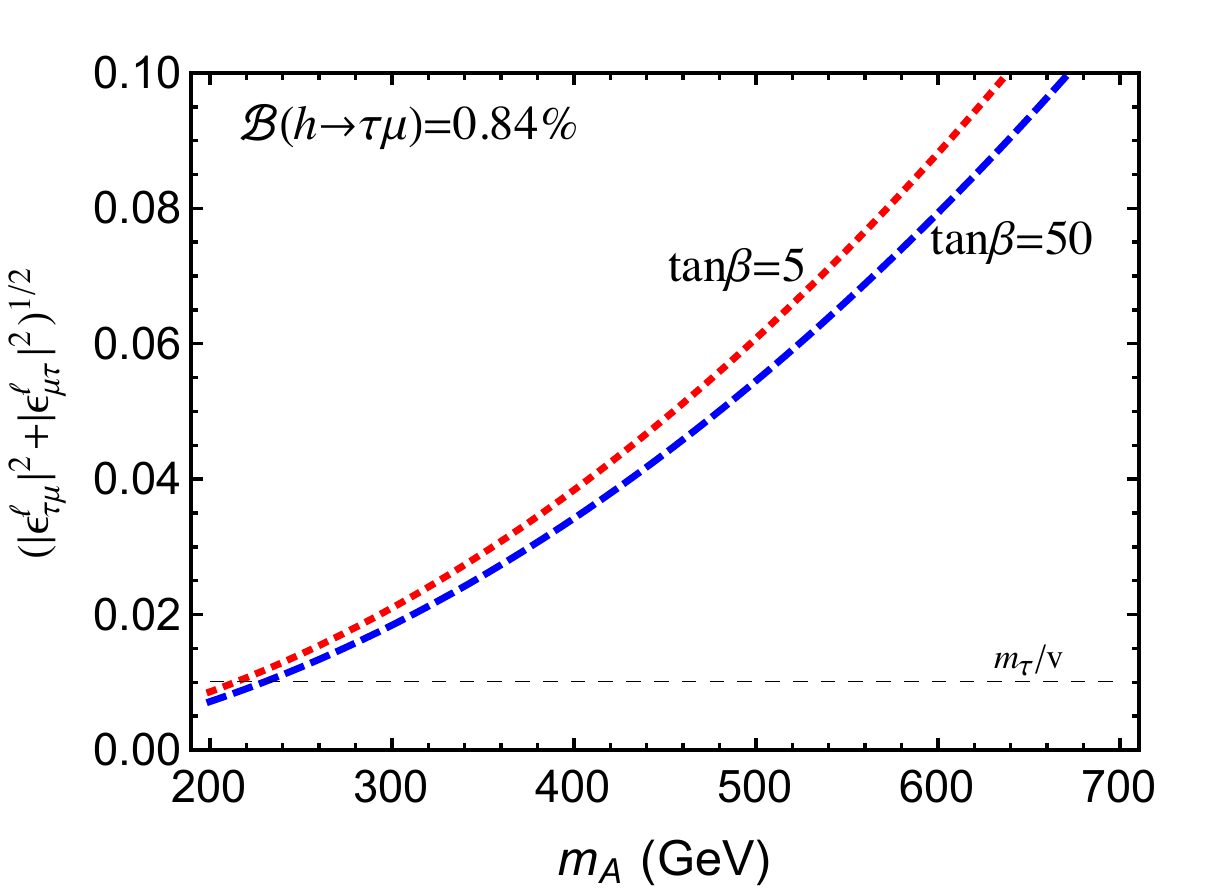}
\par\end{centering}
\caption{The values of the couplings $\epsilon_{\tau\mu}^{\ell}$ ($\epsilon_{\mu\tau}^{\ell}$) as a function of $m_A$ for which $\mc{B}(h\to\tau\mu)=0.84\%$ in type-III THDM, when all other Higgs decay rates are held SM-like. The decay exhibits small dependence on $\beta$ for large values of $\tan \beta$. See the text for details.}
\label{fig:epsVma}
\end{figure}

Using preceding relations, at the tree-level the type-III THDM contributes to the effective Higgs couplings as
\begin{equation}
  \label{eq:hLFV-2HDM}
  y_{\mu\tau\,(\tau\mu)} = \frac{\epsilon^\ell_{\mu \tau\, (\tau \mu)}}{\sqrt{2}}\,(\sin\alpha \tan \beta + \cos \alpha)\,,
\end{equation}
and we can translate Eq.~\eqref{eq:htaummu1} into a $1\,\sigma$ two-sided bound on the $\epsilon^\ell$ parameters
\begin{equation}
  \label{eq:2}
  0.0027  < |\sin \alpha \tan \beta + \cos \alpha|\, \sqrt{|\epsilon_{\tau \mu}^\ell|^2+|\epsilon_{\mu \tau}^\ell|^2} < 0.0045\,.
\end{equation}
For reference we also give the expression for the branching fraction
\begin{equation}
\label{eq:THDM-Brh}
  \mc{B}(h\to \tau \mu) = \frac{m_h}{16 \pi \Gamma_h}\,(\sin \alpha \tan\beta + \cos \alpha)^2\,\left(|\epsilon_{\mu\tau}^\ell|^2 + |\epsilon_{\tau\mu}^\ell|^2\right)\,.
\end{equation}
We show in Fig.~\ref{fig:epsVma} the points in the plane spanned by $(\vert\epsilon_{\tau\mu}^{\ell}\vert^2+\vert \epsilon_{\mu\tau}^{\ell}\vert^2)^{1/2}$ and $m_A$ for which $\mc{B}(h\to\tau\mu)=0.84\%$ when all other Higgs decay rates are held SM-like.
Interestingly, the decay exhibits small dependence on $\beta$ for large values of $\tan\beta$. Indeed, an expansion around the decoupling limit in small $\xi = m_Z^2/m_A^2$ gives
\begin{equation}
  \sin \alpha \tan\beta + \cos \alpha \simeq  -\frac{2 \tan \beta (\tan^2 \beta-1)}{(\tan^2\beta+1)^{3/2}}\xi \overset{t_\beta \to \infty}{\longrightarrow} -\frac{2 m_Z^2}{m_A^2} \,.
\end{equation}
For light  pseudo-scalar masses  ($m_A$), mixing with the heavy scalar Higgs could affect the couplings of the light Higgs to the gauge bosons. However, the overall modification factor of $hWW$ and $hZZ$ couplings relative to the SM values is given by $\sin(\beta - \alpha)$ 
and starts to differ from $1$ only at order $\mc{O}(\xi^2)$:
\begin{equation}
\label{eq:hWW}
\sin(\beta - \alpha)  \simeq  1 -  \frac{2 t_\beta^2 (t_\beta^2-1)^2}{(t_\beta^2+1)^4}\,\xi^2\,.
\end{equation}
We have checked that even for $m_A = 150\e{GeV}$ the $hWW$ coupling deviates less that $2\,\%$ from its SM value, for any value of $\tan \beta$. We conclude that the effect of modified light Higgs couplings to
gauge bosons can be safely neglected at the current precision of Higgs properties measurements. 
The more important effects of possible modifications of the light Higgs (tau) Yukawa couplings are discussed in the next subsection.

\subsection{Constraints from $\tau \to \mu \gamma$}
The coefficients $c_{L,R}$ in Eq.~\eqref{eq:Leff} are generated via one-loop 
diagrams with charged or neutral
Higgses. In the limit $m_\mu = 0$ (for
the sake of consistency also $y^{H_k}_{\mu\mu} = \epsilon_{\mu\mu}^\ell =
0$), these coefficients are~\cite{Crivellin:2013wna}
\begin{equation}
\label{eq:tmg-1loop}
\begin{split}
  c_{R}^{\textrm{({\rm 1-loop})}} &\simeq  \sum_{k=1,2,3} \frac{1}{48
    m_{H_k^0}^2} \left[y_{\mu\tau}^{H_k} y_{\tau\tau}^{H_k*}  -
    y_{\mu\tau}^{H_k} y_{\tau\tau}^{H_k} (9+6 \log
    \frac{m_\tau^2}{m_{H_k^0}^2})\right] - \frac{1}{96 m_{H^+}^2} \epsilon_{\tau\mu}^{\ell*} 
  t_\beta^2 \left(s_\beta\epsilon^\ell_{\tau\tau}- \frac{\sqrt{2}
      m_\tau}{v}\right)  \,, \\
  c_{L}^{\textrm{({\rm 1-loop})}} &\simeq  \sum_{k=1,2,3} \frac{1}{48
    m_{H_k^0}^2} \left[y_{\tau\mu}^{H_k*} y_{\tau\tau}^{H_k}  -
    y_{\tau\mu}^{H_k*} y_{\tau\tau}^{H_k*} (9+6 \log
    \frac{m_\tau^2}{m_{H_k^0}^2})\right]\,.
\end{split}  
\end{equation}
The suppression of the above result by two small Yukawa couplings  is relaxed at the two-loop level, where Barr-Zee type diagrams require only a single LFV Yukawa coupling along with $y_{tt}$. Therefore the contributions of the neutral Higgses at the two-loop level can dominate over the one-loop result.
 Here we employ the Barr-Zee contributions due to neutral Higgses calculated for $\mu \to e \gamma$ in Ref.~\cite{Chang:1993kw} and adapted to $\tau \to \mu \gamma$ in Appendix~\ref{app:BZ}. The 2-loop contributions to $\mathcal B(\tau \to \mu \gamma)$ involve exactly the same combination of $\epsilon^\ell$ elements as $\mc{B}(h \to \tau \mu)$ in Eq.~\eqref{eq:THDM-Brh}. For the charged Higgs contributions  to the best of our knowledge no complete calculation exists in the literature. A partial result with the charged Higgs contribution inducing the $H^0_k \gamma \gamma^*$ or $H^0_k \gamma Z^*$ vertices has been presented in Ref.~\cite{Hisano:2010es}. However, the diagrams with an effective $H^\pm W^\mp \gamma$ coupling via loop of $t$ and $b$ are still missing. In the following we use the known partial results to estimate the order of magnitude of the constraint. The missing contributions could potentially further strengthen the $\tau\to \mu \gamma$ bound but are not expected to substantially weaken it.

In order to explore the nontrivial correlation between the $\mc{B}(h \to \tau \mu)$ and 
$\mc{B}(\tau \to \mu \gamma)$ we sample the parameter space of $\epsilon^\ell_{\tau \mu}$, $\epsilon^\ell_{\mu \tau}$, $\epsilon^\ell_{\tau\tau}$ for a specific choice of $\tan \beta$ and $m_A$.
The ranges allowed for $\epsilon^\ell_{\tau \mu,\mu\tau}$ are required to fulfill the naturalness criterium of Eq.~\eqref{eq:lfv-nat}
\begin{equation}
  |\epsilon^\ell_{\mu \tau} \epsilon^\ell_{\tau \mu}| < \frac{m_\mu m_\tau}{v^2/2}\,(\sin\alpha \tan \beta + \cos \alpha)^{-2}  \,,
\end{equation}
 and that each of the effective couplings is within the perturbative range 
 \beq
 |\epsilon^\ell_{\mu \tau, \tau\mu}| < \sqrt{4 \pi}\,\sqrt{2}|\sin\alpha \tan \beta + \cos \alpha|^{-1}\,.
 \eeq

The 1-loop amplitude of the $\tau \to \mu \gamma$ process further depends on the diagonal $y_{\tau\tau}^h$ Yukawa coupling  which is experimentally
constrained by the searches for $h \to \tau\tau$ decays within ATLAS and CMS experiments. In fact both collaborations report strong evidence for the existence of this mode with the signal strengths (normalized to SM expectations) $\mu^{\tau\tau} = 1.43^{+0.43}_{-0.37}$ and $\mu^{\tau\tau} = 0.78 \pm 0.27$, observed by ATLAS~\cite{Aad:2015vsa} and CMS~\cite{Chatrchyan:2014nva}, respectively. In the following we use a na\"ive average of the two experimental results:
\begin{equation}
\label{eq:ytaurange}
\mu^{\tau\tau} = 1.02^{+0.21}_{-0.20} \,.
\end{equation}
In the decoupling limit (c.f. Eq.~\eqref{eq:hWW}) and with $y^h_{tt}$ SM-like, we can assume SM-like Higgs production cross sections and model the departure of $\mu^{\tau\tau}$ from 1 by an appropriate shift of $y_{\tau\tau}^h$ away from $-\sqrt{2} m_\tau/v$ (i.e. non-vanishing $\epsilon^\ell_{\tau\tau}$).

A scenario with SM-like $y^h_{\tau\tau}$ coupling corresponding to $\mu^{\tau\tau} = 1$ for fixed $\tan \beta = 10$ and $m_A = 0.3\e{TeV}$  is presented in Fig.~\ref{fig:master} by a black narrow stripe. This scenario easily passes both experimental constraints. A perfect one-to-one correspondence between the two observables is spoiled by the asymmetry between the left- and right-handed $\tau \to \mu \gamma$ amplitudes at one-loop, see Eq.~\eqref{eq:tmg-1loop}.  On the other hand, for masses $m_A$ significantly larger than $500\e{GeV}$ it is not possible to reconcile both predictions with the corresponding experimental values. 

Allowing $\epsilon_{\tau\tau}^\ell$ to vary within the experimental bounds imposed by the $h \to \tau\tau$ decays, i.e. Eq.~\eqref{eq:ytaurange}, one obtains predictions for $h\to \tau\mu$ and $\tau \to\mu\gamma$ as represented by the green band in Fig.~\ref{fig:master}. In particular, this additional freedom in $\tau \to \mu \gamma$ breaks, to some extent, the strict correlation with the $h \to \tau \mu$ rate. The remaining correlation signals the dominance of the 2-loop contributions in $\mc{B}(\tau \to \mu \gamma)$ which are independent of $\epsilon_{\tau\tau}^\ell$.

The interaction vertices that induce the LFV decay of Higgs boson to $\tau\mu$ pairs also contribute to the LFV decay of tau lepton to three muons, $\tau^-\to\mu^-\mu^+\mu^-$. The current best experimental upper limit on its branching fraction has been reported by the Belle Collaboration $\mc{B}(\tau^-\to\mu^-\mu^+\mu^-)\leq 2.1\E{-8}$ at the $90\%$ C.L. ~\cite{Hayasaka:2010np}.
This process proceeds through the tree level exchange of the neutral scalars $h,H$ and the pseudoscalar $A$. Since all the relevant amplitudes are suppressed by flavor-diagonal muon Yukawa, this constraint is not competitive with $\tau \to \mu \gamma$ in terms of sensitivity to $\epsilon^\ell_{\tau\mu}$ and $\epsilon^\ell_{\mu\tau}$.

\subsection{$h \to \tau \mu$ vs.\ $h \to \tau e$}
Here we repeat the derivation of the bound on $h \to \tau e$ described in Sec.~\ref{sec:hlfv-corr-eft} in the context of the type-III THDM model. In this case, both $\mu \to e \gamma$ and $\mu-e$ conversion receive contributions not from a single Higgs but from three neutral Higgses while the charged Higgs contributions are relatively suppressed by light lepton masses. The effective operator coefficients presented in Sec.~\ref{sec:hlfv-corr-eft} now have to be summed over the three Higgses with proper mass and effective coupling replacements. Ultimately, the observables depend on $t_\beta \equiv \tan \beta$, $m_A$, and the LFV parameters $\epsilon^\ell_{ij}$, where dependence on the first two parameters can be absorbed into an overall coefficient:
\begin{equation}
\begin{split}
  \mc{B}(\mu \to e \gamma) &\simeq \mc{B}_0^{\mu \to e \gamma} (t_\beta, m_A) \left(|\epsilon_{\mu \tau} \epsilon_{\tau e}|^2 + |\epsilon_{e \tau} y_{\tau \mu}|^2\right)\,,\qquad \mc{B}_0^{\mu \to e \gamma} (t_\beta = 10, m_A = 0.5\e{TeV}) = 2.67\,,\\
 \mathcal B(\mu\to e)_{\rm Au} &\simeq
 \mc{B}_0^{\mu e}(t_\beta, m_A)\,\left(|\epsilon_{e\tau} \epsilon_{\mu \tau}|^2 + |\epsilon_{\tau e} \epsilon_{\tau \mu}|^2\right)\,,\qquad \quad \mc{B}_0^{\mu e} (t_\beta = 10, m_A = 0.5\e{TeV}) = 0.037\,.
\end{split}
\end{equation}
Tree level LFV Higgs decays $h \to \tau \mu$ and $h \to \tau e$ are still properly described in the effective framework of Eq.~\eqref{eq:hLFV-2HDM}. In contrast with what has been found in the effective framework  (c.f. Eq.~\eqref{eq:mueconvEFT}), the sensitivity to $\epsilon^{\ell}_{ij}$ is now similar in both low energy observables. Also, the coefficients ${B}_0^{\mu \to e \gamma,\mu e}$ now have a strong dependence on $\tan \beta$ and $m_A$. Finally, Higgs decays are suppressed by a projection factor so that
\begin{equation}
\begin{split}
  \label{eq:taue-taumuTHDM}
  \mc{B}(h \to \tau \mu) \times   \mc{B}(h \to \tau e) &= \left[\frac{m_h}{16\pi\Gamma_h}\right]^2 \left(\sin \alpha \tan \beta + \cos \alpha \right)^4\left(\frac{\mc{B}(\mu \to e \gamma)}{\mc{B}_0^{\mu \to e \gamma}(t_\beta, m_A) } + \frac{\mathcal B(\mu\to e)_{\rm Au}}{\mc{B}_0^{\mu e}(t_\beta, m_A) }\right)\,.
\end{split}
\end{equation}
These  features suppress the $h \to \tau e$ process very stringently in the type-III THDM model:
\begin{equation}
\begin{split}
\mc{B} (h \to \tau e) &< 6\E{-6}\,;\qquad m_A > 0.3\e{TeV}  \,,\quad \tan\beta > 2 \,.
\end{split}
\end{equation}
The above bound is monotonically decreasing with $\tan \beta$ and $m_A$ as can be seen on Fig.~\ref{fig:htaue-htaumu}. Note that  at the projected sensitivity of the Mu2e experiment, the right hand side of Eq.~\eqref{eq:taue-taumuTHDM} would start to be dominated by the $\mu \to e \gamma$ constraint, so an improvement expected from the planned MEG upgrade (MEGII)~\cite{Baldini:2013ke} will be essential in this context, as can be seen by comparing the two bottom-most dashed contours in the plot.

\section{Vector-like leptons}
\label{sec:VLL}
In this section we discuss the possibility of generating sizable LFV Higgs couplings by mixing the SM chiral leptons with new vector-like leptons~(VLLs). Such states can appear in  the low energy spectra of grand unified theories~\cite{Shafi:1999ft, Malinsky:2007qy, Babu:2012pb,Barr:2012ma,Barr:2013dca} and are predicted to exist in composite Higgs scenarios with partial compositeness~\cite{Contino:2006nn, Agashe:2006iy, Agashe:2009tu, Redi:2013pga}. In this setting the chiral leptons obtain additional couplings to the Higgs by mixing with the heavy vector-like leptons.

We first consider the addition of VLLs in a single vectorial representation of the SM gauge group $SU(3)_c \times SU(2)_L\times U(1)_Y$, i.e. $(1,2)_{1/2} \oplus (1,2)_{-1/2}$ or $(1,1)_{1} \oplus (1,1)_{-1}$. In this case one can prove~\cite{Fajfer:2013wca} that LFV Higgs couplings are directly related to LFV $Z$-boson couplings 
\beq
\mathcal L_{\rm LFV}^{Z} = \frac{g}{2c_W} \left(  X_{ij} \bar\ell^i_L \gamma^\mu \ell^j_L - Y_{ij} \bar\ell^i_R \gamma^\mu \ell^j_R \right) Z_\mu\,,
\eeq
where $g\simeq 0.65$ and $c_W\equiv \cos \theta_W \simeq 0.88$ are the weak isospin gauge coupling and the cosine of the weak mixing angle, respectively. In particular one obtains a relation $y_{ij} = (m_j/v) (X_{ij} - Y_{ij})$\,. Severe constraints on $X_{\tau\mu,\mu\tau},Y_{\tau\mu,\mu\tau} \lesssim 10^{-3}$ from searches for $\tau \to 3 \mu$~\cite{Dorsner:2014wva} then preclude any sizable contributions to $\mathcal B(h\to\tau\mu)$\,.\footnote{A similar conclusion can be drawn in the case of weak-isospin triplet Majorana fermions, appearing in Type-III see-saw models of neutrino masses~\cite{Kamenik:2009cb}.} De-correlating LFV Higgs and $Z$ boson couplings requires the introduction of VLLs in both SM gauge representations ($\Psi^E, \Psi^L$) mixing with chiral leptons 
\bea
- \mathcal L_{VLL} &=&  \lambda_\Psi \bar \Psi^E H (1-\gamma_5)\Psi^L + \tilde \lambda_\Psi \bar \Psi^E H (1+\gamma_5) \Psi^L \nonumber\\
&&+ M_{\Psi} \left( \lambda_e \bar E \Psi^E + \lambda_l \bar L \Psi^L + C_L \bar \Psi^L \Psi^L + C_R \bar \Psi^E \Psi^E \right) + \rm h.c.\,.
\label{eq:VLLmix}
\eea
Integrating out the vector-like states one finds the LFV Higgs and $Z$ boson couplings can now be completely de-correlated in the limit where the direct couplings of chiral fermions to the Higgs vanish, and the SM leptons obtain their masses exclusively through mixing with VLLs in Eq.~\eqref{eq:VLLmix}. In particular the lepton flavor off-diagonal Higgs Yukawas in Eq.~\eqref{eq:ye} are then of the form~\cite{Falkowski:2013jya}
\begin{equation}
\epsilon = \frac{8 v^2}{M_\Psi^2}\, \lambda_l C_L^{-1} \lambda_\Psi C_R ^{-1} \tilde \lambda_\Psi C_L^{-1} \lambda_\Psi C_R^{-1} \lambda_e\,.  
\end{equation}
All the LFV phenomena are driven by this contribution and one can derive a one-to-one correlation between tree-level Higgs decay $h \to \tau \mu$ and the 1-loop radiative decay $\tau \to \mu \gamma$~\cite{Falkowski:2013jya}
\begin{equation}
\frac{\mc{B}(h \to \tau \mu)}{\mc{B}(\tau \to \mu \gamma)} = \frac{4\pi}{3\alpha}\, \frac{\mc{B}(h \to \tau^+ \tau^-)_\mrm{SM}}{\mc{B}(\tau \to \mu \bar \nu \nu)_{\rm SM}} \approx 2\E{2}\,.
\end{equation}
The 1-loop suppression factor of $\tau \to \mu \gamma$ is not small enough to evade the experimental upper bound and at the same time accommodate the $h \to \tau \mu$ at the percent level, as illustrated by a pink-dotted line in Fig.~\ref{fig:master}.

\section{Scalar Leptoquarks}
\label{sec:LQ}
\begin{figure}[t]
\begin{centering}
\includegraphics[scale=0.55]{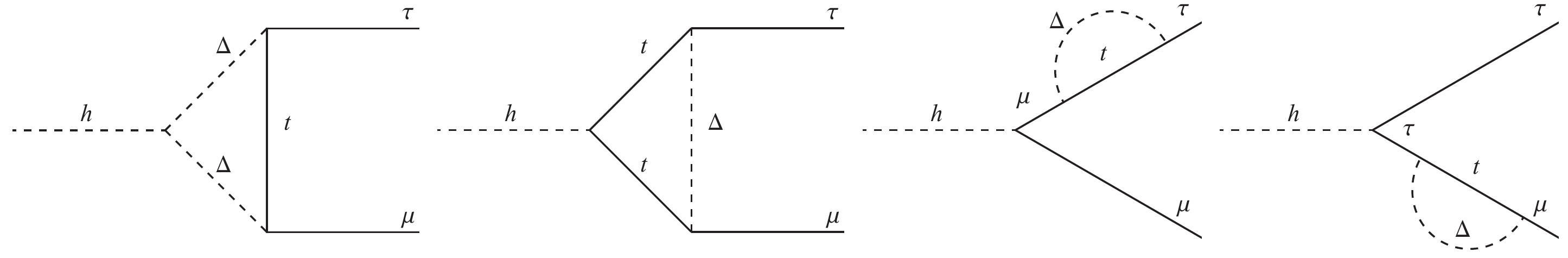}
\end{centering}
\caption{Feynman diagrams for LQ contribution to $h\to \tau \mu$.}
\label{fig:LQdiag}
\end{figure}
A scalar leptoquark state (LQ) can induce $h \to \tau \mu$ decay via quark-LQ penguin diagrams. This type of diagrams requires a helicity flip on one of the fermion lines. One therefore expects suppressed amplitude when all the fermions are substantially lighter than $v$. It turns out that size of LQ Yukawa couplings must then be well beyond the perturbative regime to result in $\mathcal{B}(h \to \tau \mu)$ at the percent level. In this respect, only LQ states that couple to charged leptons and top quark are suitable candidates that will be studied in what follows. Inspection of helicity structure of the diagrams reveals that both left and right chiralities of leptons and top quark have to couple to the LQ state.

The physical Higgs can couple to the scalar $\Delta$ or to the top quark as shown in the first two diagrams in Fig.~\ref{fig:LQdiag}.  While the strength of the coupling relevant for the latter process is fixed by the top Yukawa, the former process depends on an unknown $h \mrm{LQ} \mrm{LQ}$ coupling, $\lambda v$, that originates from the marginal ``Higgs portal" operator, 
\begin{equation} 
\label{eq:portSing} \mathcal{L} \ni-\lambda H^{\dagger}H\Delta^{\dagger}\Delta\,.  
\end{equation} 
Here $\Delta$ is the scalar LQ and $H$ is the SM Higgs doublet.

\subsection{The $\Delta_1=(\bf{3},\bf{1},-1/3)$ case}
The Yukawa couplings of $\Delta_1$ are given by the following Lagrangian
\begin{equation}
\mathcal{L}_\mathrm{\Delta_1}=y^L_{ij}\bar{Q}_{}^{i,a}\Delta_1\epsilon^{ab}L_{}^{C\,j,b}+y^R_{ij}\bar{U}_{}^{i}\Delta_1 E_{}^{C\,j}+\textrm{h.c.}\,,
\end{equation}
where $Q^i=(u^i_L,d^i_L)^T$ and $U^i=u^i_R$ are the quark weak doublets and up-type singlets, respectively.  We explicitly show flavor indices $i,j=1,2,3$, and $SU(2)$ indices $a,b=1,2$, with $\epsilon_{12} = 1$. Also, here $y^L_{ij}$ and $y^R_{ij}$ are elements of arbitrary complex $3 \times 3$ Yukawa coupling matrices. After expanding the $SU(2)$ indices, we obtain
\begin{equation}
\label{eq:main_1}
\mathcal{L}_\mathrm{\Delta_1}=y^L_{ij}\bar{u}_{L}^{i} \ell_{L}^{C\,j} \Delta_1-(V_\mathrm{CKM}^\dagger y^L V_\mathrm{PMNS})_{ij} \bar{d}_{L}^{i} \nu_{L}^{C\,j} \Delta_1+ y^R _{ij}\bar{u}_{R}^{i}\ell_{R}^{C\,j} \Delta_1 +\textrm{h.c.},
\end{equation}
where $V_\mathrm{CKM}$ and $V_\mathrm{PMNS}$ represent Cabibbo-Kobayashi-Maskawa and Pontecorvo-Maki-Nakagawa-Sakata mixing matrices, respectively. All fields in Eq.~\eqref{eq:main_1} are specified in the mass eigenstate basis. The Wilson coefficients of the $h \to \tau \mu$ effective Lagrangian~\eqref{eq:Brhtm} are obtained after summing the diagrams in Fig.~\ref{fig:LQdiag}
\begin{align}
y_{\tau\mu \,(\mu\tau)}  &= -\frac{N_c}{16\pi^2} \frac{m_t}{v} g_1(\lambda, m_{\Delta_1}) \,y^{R}_{t \mu} y^{L*}_{t \tau} \,\,\left (y^{R}_{t \tau} y^{L*}_{t \mu}\right)\,,
\end{align}
where $N_c =3$ is the number of colors.
The relevant loop function further depends on the portal coupling $\lambda$,
\begin{equation}
\begin{split}
g_1(\lambda,m_{\Delta}) &=(m_{\Delta}^{2}+m_{t}^{2})C_{0}(0,0,m_{h}^{2},m_{t}^{2},m_{\Delta}^{2},m_{t}^{2})+B_{0}(m_{h}^{2},m_{t}^{2},m_{t}^{2})-B_{0}(0,m_{t}^{2},m_{\Delta}^{2})\\
&\phantom{=}+\lambda v^{2}C_{0}(0,0,m_{h}^{2},m_{\Delta}^{2},m_{t}^{2},m_{\Delta}^{2})\,,
\end{split}
\end{equation}
given in terms of Passarino-Veltman functions $B_0$ and $C_0$\footnote{We disagree with the result of Ref.~\cite{Cheung:2015yga} where the authors employed a subtraction at kinematical point $m_h = 0$ to render the $h\to \tau \mu$ amplitude finite.}.
The  $h \to \tau \mu$ decay width in the $\Delta_1$ LQ scenario is then given by
\begin{equation}
\label{eq:LQ1-BR}
\Gamma(h\to\tau\mu)=\frac{9 m_{h}m_{t}^{2}}{2^{13}\pi^{5}v^{2}}\left(\left|y_{t\mu}^{L}y_{t\tau}^{R}\right|^{2}+\left|y_{t\tau}^{L}y_{t\mu}^{R}\right|^{2}\right)\left|g_1(\lambda,m_{\Delta_1})\right|^{2}\,.
\end{equation}

The state $\Delta_1$ also contributes to the $\tau \to \mu \gamma$ through leptoquark Yukawas present in $h \to \tau \mu$. The Wilson coefficients of effective Lagrangian introduced in Eq.~\eqref{eq:Leff} are
\begin{equation}
c_{L(R)}  = -\frac{N_c}{24} \frac{m_t}{m_\tau m_{\Delta_1}^2} h_1(x_t)\,y^{R*}_{t \mu} y^{L*}_{t \tau}\,(y^{R}_{t \tau} y^{L}_{t \mu}),\qquad h_1(x) = \frac{7-8x +x^2+(4+2x)\ln x}{(1-x)^3}\,,
\label{eq:48}
\end{equation}
with $x_t = m_t^2/m_{\Delta_1}^2$, in agreement with~\cite{Lavoura:2003xp}. The $\tau \to \mu \gamma$ branching fraction in presence of $\Delta_1$ depends on identical Yukawa combination as the
$h \to \tau \mu$ decay rate:
\begin{equation}
 \mathcal{B}(\tau \to \mu \gamma) = \frac{\alpha m_\tau^3}{2^{12} \pi^4 \Gamma_\tau}\,\frac{m_t^2}{m_{\Delta_1}^4} \, h_1(x_t)^2 \, \left(\left|y_{t\mu}^{L}y_{t\tau}^{R}\right|^{2}+\left|y_{t\tau}^{L}y_{t\mu}^{R}\right|^{2}\right)\,.
\end{equation}

Additional constraints on the model parameters can also come from the measurements of $(g-2)_\mu$ and the electroweak precision observables, i.e. $\mathcal B(Z\to b\bar b)$. The $\Delta_1$ contribution to the former is dominated by the coupling combination $y^L_{t\mu} y_{t\mu}^R$ which can always be suppressed. For the $Z \to b\bar b$ the presence of $\Delta_1$ will modify the effective coupling $\delta g_L^b$, defined as:
\begin{equation}
\label{eq:Zbbdef}
\mc{L}_{Zb\bar b} = \frac{g}{\cos \theta_W} Z^\mu 
    \bar b \gamma_\mu \left[(g_L^b + \delta g_L^b) P_L +  (g_R^b +
      \delta g_R^b) P_R \right] b\,.
\end{equation}
Here $g$ is the $SU(2)_L$ gauge coupling, $\theta_W$ is the Weinberg angle. $g_{L,R}^b$ are the SM effective couplings including the electroweak radiative corrections, while $\delta g_{L,R}^b$ parameterize contributions that originate from beyond the SM. The LQ state $\Delta_1$ contributes
to $\delta g_L^b$ through loops with neutrinos,
\begin{equation}
  \label{eq:dgLb1}
  \delta g_L^b = \frac{|V_{tb}|^2 \left(|y^L_{t \tau}|^2 + |y^L_{t \mu}|^2\right)}{288 \pi^2} \left[ -\frac{3\log x+1 + 3 i \pi}{x} + \frac{\sin^2\theta_W}{3x} \right]\,,
\end{equation}
where $x = m^2_{\Delta_1}/m_Z^2 \gg 1$.
The constraint on LQ Yukawas follows from combining Eq.~\eqref{eq:dgLb1} with the results of the recent fit to the 
$Z$-pole observables~\cite{Fajfer:2013wca}:
\begin{equation}
\label{eq:Zpole}
\re[\delta g_{L}^b] = (2 \pm 2)\E{-3}\,,\qquad \re[\delta g_{R}^b] = (15 \pm 6)\E{-3}\,.
\end{equation}
The $Z$-pole measurements are sensitive to interference terms between $g_{L,R}^b$ and $\delta g_{L,R}^b$, where $g_{L,R}^b$ are approximately real numbers,
and therefore only real parts of $\delta g_{L,R}^b$ give linear shift in $g_{L,R}^b$. According to Ref.~\cite{Fajfer:2013wca} the SM point corresponds to $\chi^2_\mrm{SM} = 7.5$, where $0$ is defined as the fit with free $\delta g_{L,R}^b$. The presence of LQ will push $\delta g_L^b$ in the negative direction. At $1\sigma$ we insist on $\chi^2 < \chi^2_\mrm{SM} + 1$ that corresponds to a lower limit $\Delta g_L^b > -1.2\E{-4}$. An approximate constraint on the Yukawas is finally obtained,
\begin{equation}
   \sqrt{|y^L_{t \tau}|^2 + |y^L_{t \mu}|^2} \lesssim 1.30 \,\frac{m_{\Delta_1}}{\mrm{TeV}} + 0.34\,,\qquad (\textrm{valid for } 0.5 \e{TeV} < m_{\Delta_1} < 2.0\e{TeV})\,.
\end{equation}
Clearly, this constraint does not preclude the Yukawa couplings combination entering $\mc{B}(h \to \tau \mu)$ from being large.

\begin{figure}
\begin{centering}
\includegraphics[scale=0.6]{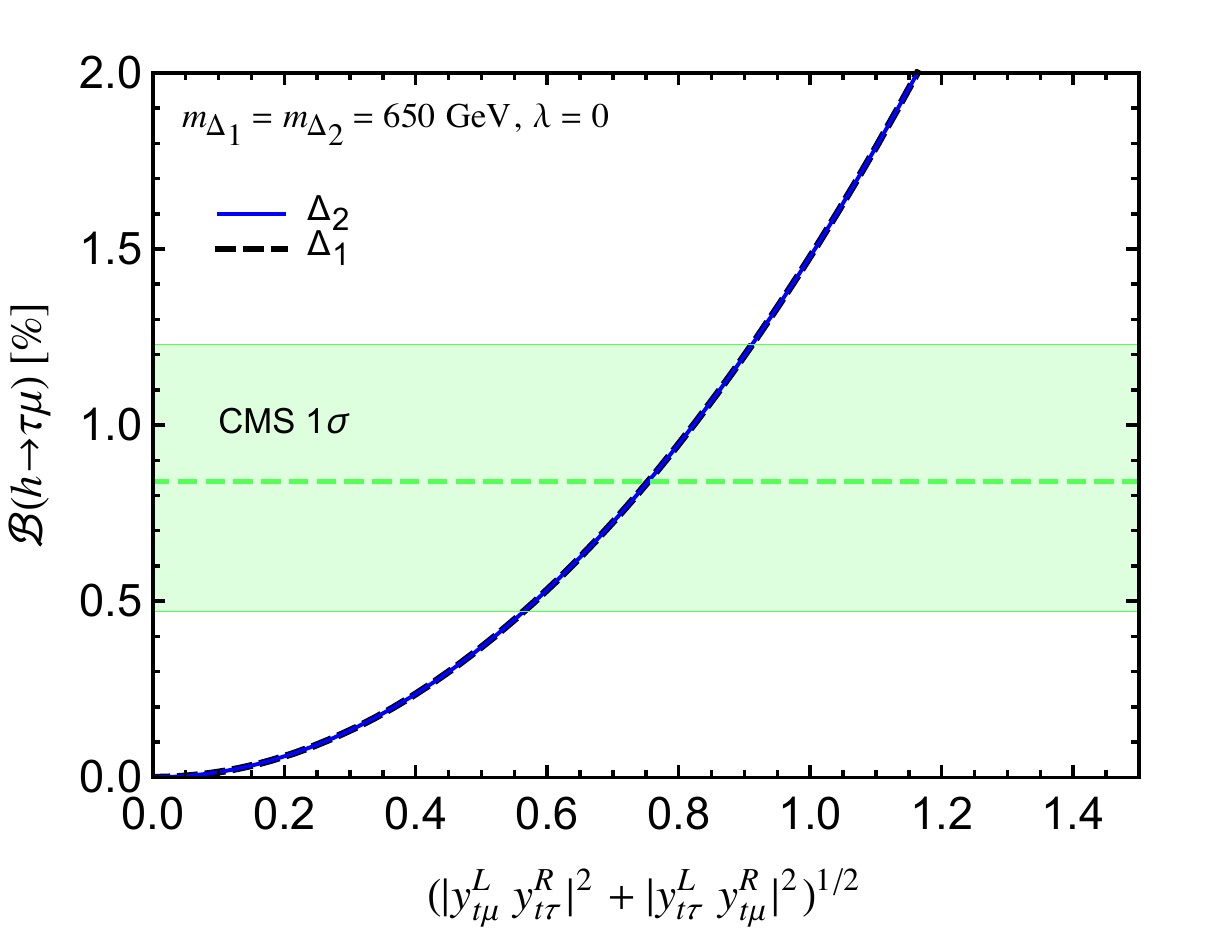} \;
\includegraphics[scale=0.63]{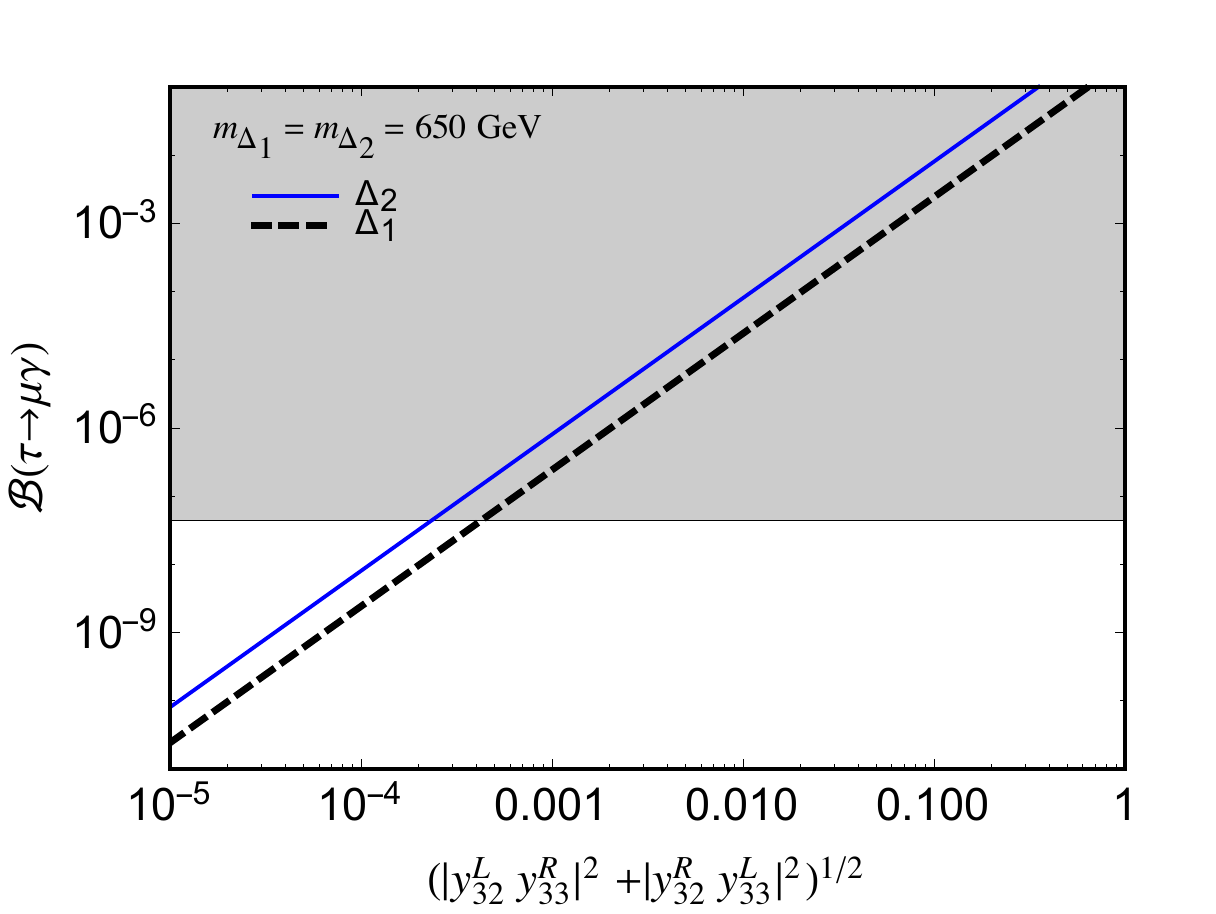}
\par\end{centering}
\caption{(Left-hand side panel) Higgs branching ratio to $\tau\mu$ final state in the presence of scalar leptoquarks $\Delta_1=(\bf{3},\bf{1},-1/3)$ and $\Delta_2=(\bf{3},\bf{2},7/6)$ with Higgs portal coupling $\lambda$ set to zero. (Right-hand side panel) $\tau\to\mu\gamma$ constraints. The $\Delta_1$ ($\Delta_2$) case is rendered in dashed (solid) line. The following transformation needs to be applied when going from the $\Delta_1$  to the $\Delta_2$ case: $y_{ij}^{L} \rightarrow y_{ji}^{L}$.}
\label{fig:-plot-2}
\end{figure}

In the left-hand side panel of Fig.~\ref{fig:-plot-2}
we show the dependence of $\mathcal{B}(h\to\tau\mu)$ on the leptoquark Yukawa couplings 
assuming the total Higgs decay width to be SM-like. Here, we turn off the Higgs portal coupling.
We set the mass of $\Delta_1$ LQ to $650$\,GeV, consistent with current direct search limits at the LHC~\cite{ATLAS:2013oea,CMS-PAS-EXO-13-010}. In the same panel of Fig.~\ref{fig:-plot-2}, we show the CMS 
measurement reported in~\cite{Khachatryan:2015kon}. The $\mathcal{B}(\tau\to \mu \gamma)$ dependence on the leptoquark Yukawa couplings for $\Delta_1$ case is shown in the right-hand side panel of Fig.~\ref{fig:-plot-2} (dashed line), again for $m_{\Delta_1} = 650\e{GeV}$. 

\subsection{Impact of the Higgs portal and its constraints}
\begin{figure}
\begin{centering}
\includegraphics[scale=0.65]{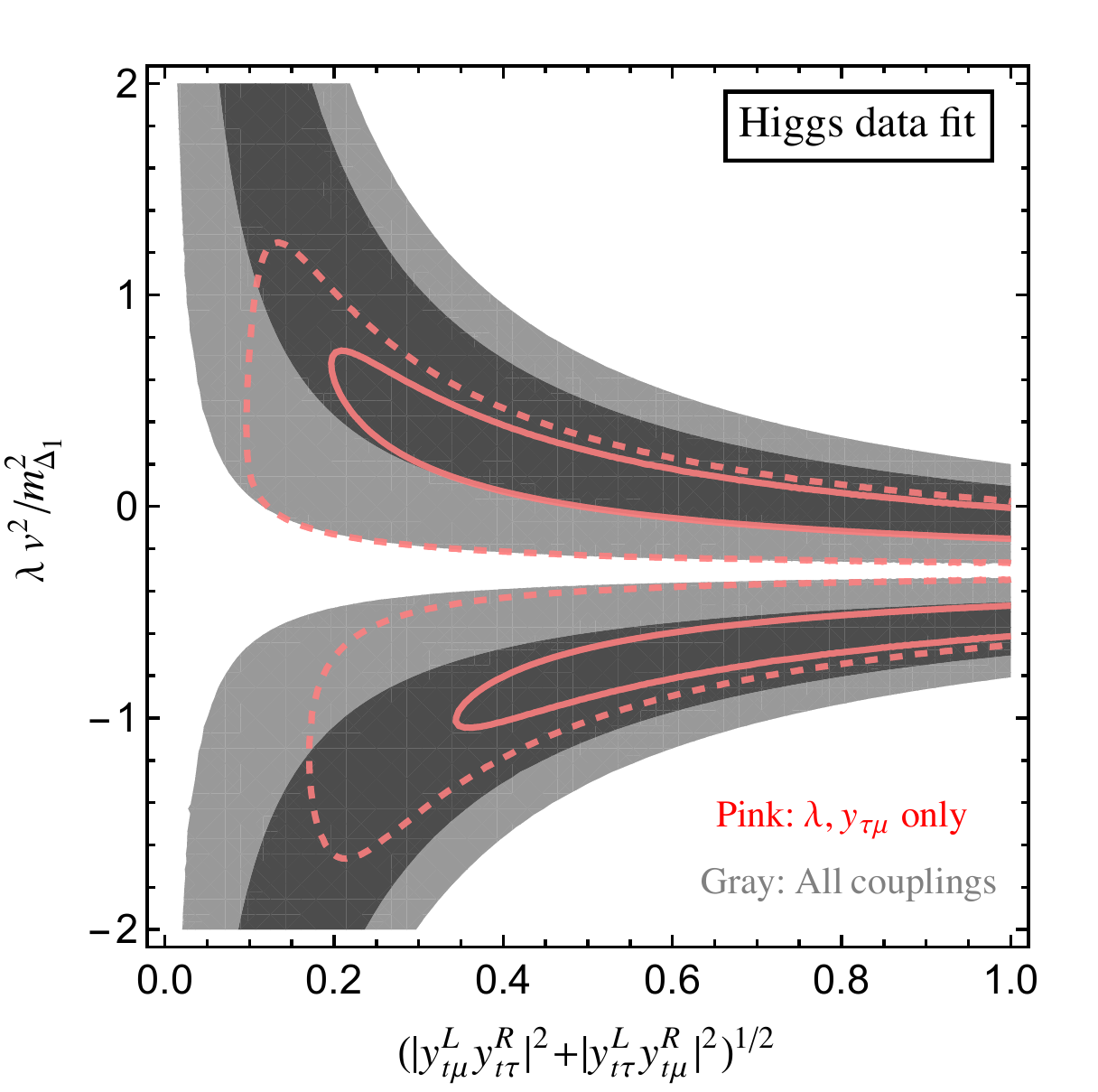}
\par\end{centering}
\caption{Fit to the latest available LHC Higgs data including the CMS $h\to\tau\mu$ measurement.}
\label{fig:-plot-3}
\end{figure}

Next we study the impact of a non-zero Higgs portal coupling in the scenario with $\Delta_1$. As an example, for $m_{\Delta_1}=650$\,GeV the
loop function dependence on the portal coupling is
\begin{equation}
g_1(\lambda, m_{\Delta_1} = 650\e{GeV})=-(0.26+0.12 \lambda)\,.
\end{equation}
Thus, a positive large $\lambda$ could in principle relax the leptoquark Yukawa couplings and
yield sizable $h\to \tau \mu$ rates without violating the $\tau \to \mu \gamma$ constraint.
However, the Higgs portal coupling also induces corrections to the $h\to\gamma\gamma$ decay
and to gluon-gluon fusion (ggF) induced Higgs production with the leptoquark running in the triangular loop.
The modified ggF production 
cross section normalized to its SM value is
\begin{equation}
\frac{\sigma_{ggF}}{\sigma_{ggF}^{SM}}=|\hat{c}_g|^2,\;\; \textrm{where}\;\;
\hat{c}_g=1+0.24  \frac{\lambda v^{2}}{m_{\Delta}^{2}}N_{\Delta{^i}} C(r_{\Delta}).
\label{eq:ggmodif}
\end{equation}
Here, $N_{\Delta^i}$ is the number of $\Delta^i$ components in the weak multiplet $\Delta$. 
$C(r_{\Delta})$ is the index of color representation $r_{\Delta}$ of  $\Delta$ and for the 
triplet ($C(\mathbf 3)=1/2$). We consider heavy enough colored scalars such that 
the loop function is in the decoupling limit. Similarly, the modified $h\to \gamma \gamma$ 
decay width, normalized to its SM value, is given by
\begin{equation}
\frac{\Gamma_{h\to\gamma\gamma}}{\Gamma_{h\to\gamma\gamma}^{SM}}=|\hat{c}_\gamma|^2,\;\; \textrm{where}\;\;
\hat{c}_\gamma=1-0.025   \frac{\lambda v^{2}}{m_{\Delta}^{2}}  d(r_{\Delta}) \underset{i}{\sum} Q_{\Delta^i}^{2} .
\label{eq:gagamodif}
\end{equation}
The sum in Eq.~\eqref{eq:gagamodif} runs over all weak components of the $SU(2)_L$ multiplet. $d(r_{\Delta})$ and  $Q_{\Delta^i}$ 
are the dimension of the color representation of $\Delta$ and the electric charges of weak $\Delta^i$ components, respectively.
We fit the latest LHC Higgs data (including the CMS signal of $h\to\tau\mu$) taking $\lambda$ and leptoquark Yukawa couplings as free parameters. The result for the $\Delta_1$ LQ model is shown in Fig.~\ref{fig:-plot-3}. The preferred regions at $1\,\sigma$ and $2\,\sigma$ are rendered in solid pink and dashed pink, respectively. If besides the $\lambda$ and leptoquark Yukawa couplings, we also allow other Higgs couplings to vary (see Appendix~\ref{sec:appHiggsFit}) we instead get the dark and light grey regions at $1\,\sigma$ and $2\,\sigma$ levels, respectively.

The correlation between the $h\to \tau \mu$ and $\tau \to \mu \gamma$ branching ratios in  presence of the $(3,1,-1/3)$ leptoquark state for $|\lambda| < 1$ and $m_{\Delta_1} > 600\e{GeV}$ is depicted in Fig.~\ref{fig:master} with a pink stripe.

\subsection{The $\Delta_2=(\bf{3},\bf{2},7/6)$ case}
The Yukawa couplings of the $\Delta_2$ leptoquark to  SM fermions are
\begin{equation}
\label{eq:main_2}
 \mathcal{L}_\mathrm{\Delta_2} = y^L_{ij}\bar{E}_{}^{i} \Delta_{2}^{a\,*}Q_{}^{j,a} -y^R_{ij}\bar{U}_{}^{i} \Delta_{2}^{a}\epsilon^{ab}L_{}^{j,b}+\textrm{h.c.},
\end{equation}
where we explicitly show the flavor indices $i,j=1,2,3$, and $SU(2)$ indices $a,b=1,2$. $y^L$ and $y^R$ in Eq.~\eqref{eq:main_2} are arbitrary complex $3 \times 3$ Yukawa matrices. 
In the mass eigenstate basis we have 
\begin{equation}
  \label{eq:L23}
\mathcal{L}_\mathrm{\Delta_2}= y^L_{ij} \bar{\ell}^i_R d^j_L \Delta_2^{2/3\,*} +(y^L V_\mathrm{CKM}^\dagger)_{ij} \bar{\ell}^i_R 
  u^j_L \Delta_2^{5/3\,*}+  (y^R
  V_\mathrm{PMNS})_{ij} \bar{u}^i_R 
  \nu^j_L \Delta_2^{2/3}  - y^R_{ij} \bar{u}^i_R \ell^j_L \Delta_2^{5/3} + \textrm{h.c.},
\end{equation}
where a superscript on $\Delta_2$ denotes the electric charge of a given $SU(2)$ doublet component. Note that only the $\Delta_2^{5/3}$ state couples to charged leptons and the top quark and thus contributes to $h \to \tau \mu$ decay via virtual top quark. Due to the weak-doublet nature, this state has an additional Higgs portal operator of the form $(H^{\dagger}\Delta)\, (\Delta^{\dagger} H)$ that does not induce $\Delta_2^{5/3}$ coupling to the physical Higgs field. The effective $h\tau\mu$ couplings are
\begin{equation}
y_{\tau \mu\,(\mu \tau)} = \frac{N_c}{16\pi^2} \frac{m_t}{v} g_1(\lambda, m_{\Delta_2})\,y^{L*}_{\mu t} y^{R*}_{t \tau}\,\,\left(y^{L*}_{\tau t} y^{R*}_{t \mu}\right)\,,
\end{equation}
where the loop function $g_1$ has been introduced in the previous section.
The $h \to \tau \mu$ decay rate is then
\begin{equation}
  \Gamma(h \to \tau \mu) = \frac{9 m_h m_t^2}{2^{13} \pi^5 v^2} |g_{1}(\lambda, m_{\Delta_2})|^2 \left(
|y_{\mu t}^L y_{t\tau}^R|^2 +
|y_{\tau t}^L y_{t\mu}^R|^2
 \right)\,.
\end{equation}
Allowed values for the Higgs portal coupling $\lambda$ can be inferred from a global fit to the Higgs data as has been done for the portal coupling of the $(3,1,-1/3)$ state. (See Fig.~\ref{fig:-plot-3}.)
We do not attempt to repeat the same procedure for the state $(3,2,7/6)$ since we do not expect a drastic change in the allowed range of $\lambda$. 

The Wilson coefficients for $\tau \to \mu \gamma$ are again proportional to the couplings responsible for $h \to \tau \mu$:
\begin{equation}
\label{eq:tmg-d2}
c_{L\,(R)} = -\frac{N_c}{24} \frac{m_t}{m_\tau m_{\Delta_2}^2}\,h_2(x_t)\,y_{t \tau}^R y_{\mu t}^L\, \left(y_{t \mu}^{R*} y_{\tau t}^{L*}\right)\,, \qquad  h_2(x) = \frac{1-8x + 7x^2+(4-10x)\ln x}{(1-x)^3}\,,  
\end{equation}
with $x_t = m_t^2 / m_{\Delta_2}^2$ and agree with the formulas presented in~\cite{Lavoura:2003xp}. Finally, the corresponding branching fraction is given by
\begin{equation}
  \label{eq:LQ76-tau-mug}
  \mathcal{B} (\tau \to \mu \gamma) = \frac{\alpha m_\tau^3}{2^{12} \pi^4 \Gamma_\tau} \frac{m_t^2}{m_\Delta^4}\,h_2(x_t)^2\, \left(|y_{t \tau}^R y_{\mu t}^L|^2 + |y_{t \mu}^R y_{\tau t}^L|^2\right)\,.
\end{equation}

As in the case of the $\Delta_1$ leptoquark, contributions to the muon anomalous magnetic moment $(g-2)_\mu$ are proportional to $y^R_{t\mu} y^L_{\mu t}$ and can be always suppressed. On the other hand, contributions of $\Delta_2$ enter the $Z \to b \bar b$ via positive contributions to $\delta g_L^b$ that can actually help in easing the tension with the experimental data~\cite{Dorsner:2013tla}. Most importantly, these observables do not constrain large possible $\Delta_1$ effects in  $h\to \tau\mu$\,.

In the left-hand (right-hand) side panel of Fig.~\ref{fig:-plot-2} we show the $\mathcal{B}(h\to\tau\mu)$ ($\mathcal{B}(\tau\to \mu \gamma)$) dependence on the leptoquark Yukawa couplings 
for the $\Delta_2$ LQ case with the Higgs portal coupling turned off and taking $m_{\Delta_2} = 650\e{GeV}$. 
Also in this leptoquark scenario the bound on $\mc{B}(\tau \to \mu \gamma)$ excludes sizable $\mc{B}(h \to \tau \mu)$ due to the strict correlation between the two observables. See the orange stripe in Fig.~\ref{fig:master}, where the portal coupling is restricted to $|\lambda| < 1$.

\subsection{Fine-tuning solution}
\label{sec:finetune}

\begin{figure}
\begin{centering}
\includegraphics[scale=0.75]{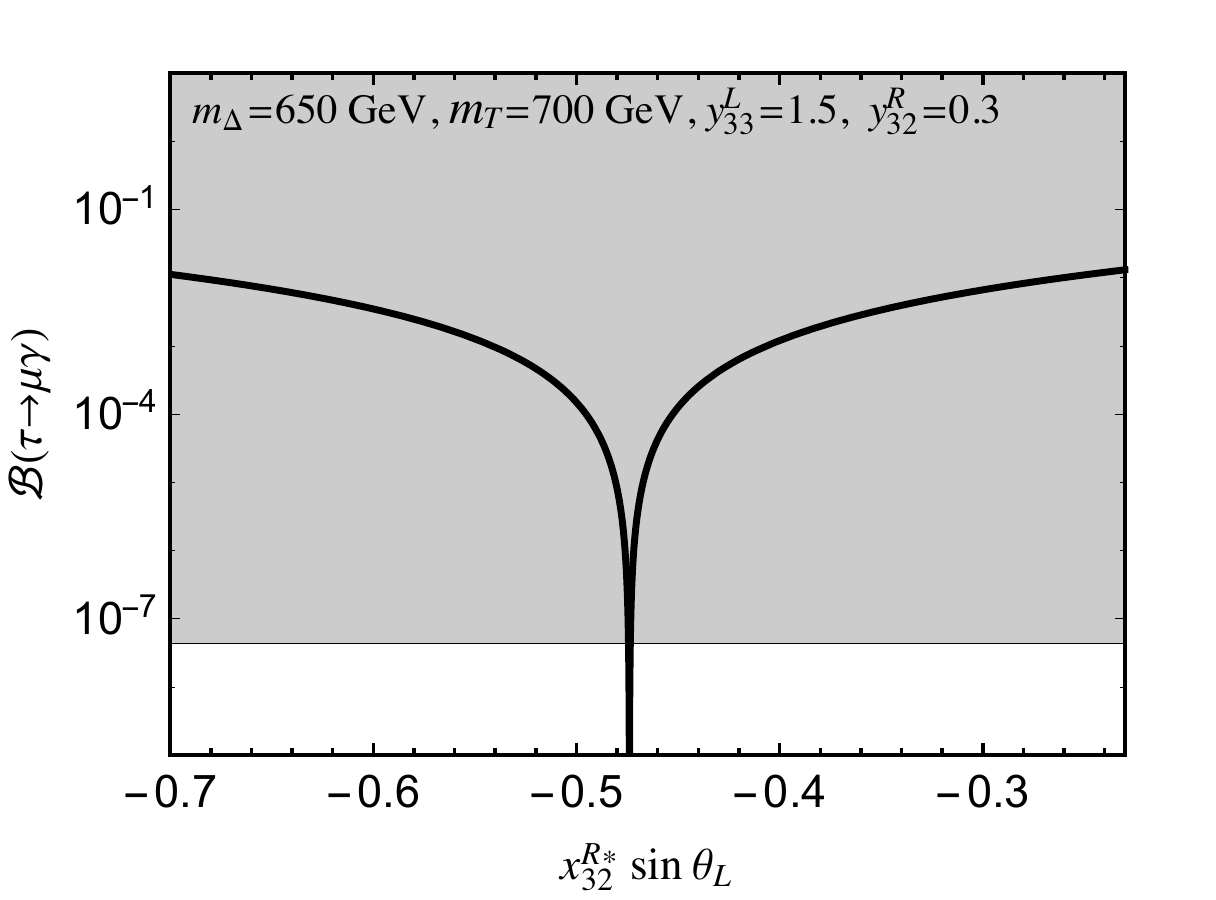}
\par\end{centering}
\caption{ Fine-tuning $\tau\to\mu\gamma$ to zero while keeping $h\to \tau \mu$ at best fit point within the LQ model with a vector-like top quark partner. See text for details.}
\label{fig:-plot-FT}
\end{figure} 

In this section we give an example of a phenomenologically viable model with fine-tuning. We add to the SM a scalar leptoquark $\Delta_1=(\bf{3},\bf{1},-1/3)$ and a vector-like top quark partner $T'_L, T'_R=(\bf{3},\bf{1},2/3)$. In the following, the top partner will mix with the SM top quark. In full generality, the relevant Lagrangian is given by~\cite{Fajfer:2013wca}
\begin{equation}
-\mathcal{L} \supset y_t \bar q'_{3L}\tilde H t'_R+y_T \bar q'_{3L}\tilde H T'_R+M_T \bar T'_L T'_R+\textrm{h.c.}~.
\end{equation}
After EWSB, the mass eigenstate fields $t$ and $T$ are obtained after rotating $t'$ and $T'$ with left- and right-handed mixing angles, $\theta_L$ and  $\theta_R$. In the  phenomenologically viable limit of a heavy top partner (and small mixing), the following relations hold~\cite{Fajfer:2013wca}
\begin{equation}
m_t \approx y_t v / \sqrt{2} ~, \; \; m_T \approx M_T~, \;\; \sin \theta_L  \approx \frac{m_t y_T}{m_T y_t}~, \;\; \sin \theta_R \approx \frac{m_t}{m_T} \sin\theta_L~. 
\end{equation}
More precisely, the direct LHC searches for a singlet top partner set a lower bound on $m_T$, while the electroweak precision observable ($\rho$ parameter) constraints the mixing angle. In what follows, we set $m_T=700$~GeV and $\sin \theta_L=0.2$ which are the borderline values compatible with the constraints~\cite{Fajfer:2013wca}.
The relevant LQ interactions are
\begin{equation}
\mathcal{L} \supset y^L_{3j} \bar{q'}_{3L}^{a} \Delta_1\epsilon^{ab}L_{}^{C\,j,b}+y^R_{3j} \bar t'_{R} \Delta_1 E_{}^{C\,j}+x^R_{3j}\bar T'_{R} \Delta_1 E_{}^{C\,j}+\textrm{h.c.}\,.
\end{equation}
After rotating to the mass basis
\begin{equation}
\footnotesize
\mathcal{L} \supset y^L_{3j} (\cos \theta_L \bar{t}_{L}+\sin \theta_L \bar{T}_{L}) \ell_{L}^{C\,j} \Delta_1+y^R_{3j} (\cos \theta_R \bar t_{R}+\sin \theta_R \bar T_R)  \ell_{R}^{C\,j} \Delta_1+x^R_{3j}(\cos \theta_R \bar T_{R}-\sin \theta_R \bar t_R)  \ell_{R}^{C\,j} \Delta_1+\textrm{h.c.}\,.
\end{equation}
Now, $\tau\to\mu\gamma$ amplitude will receive contributions both from $t$ and $T$ at one loop level.
Assuming non-zero $y_{33}^L$, $y_{32}^R$ and $ x_{32}^{R}$, neglecting the right-handed mixing and linearising in $\sin \theta_L$, we find the branching ratio
\begin{equation}
\mathcal{B}(\tau\to\mu\gamma)= \frac{\alpha_\mrm{EM} m_\tau^3 |y_{33}^{L}|^2}{2^{12} \pi^4  \Gamma_\tau m_{\Delta_{1}}^4} 
\left|y_{32}^{R*}\, m_t h_1(m_t^2/m_{\Delta_{1}}^2) +   x_{32}^{R*} \sin \theta_L \, m_T h_1(m_T^2/m_{\Delta_{1}}^2) \right|^2\,,
\end{equation}
where $h_1(x)$ is defined in Eq.~\eqref{eq:48}. As a numerical example, we take $m_{\Delta_1}=650$\,GeV, $m_T=700$~GeV and find the condition for the complete cancellation in $\mathcal{B}(\tau\to\mu\gamma)$ to be $y_{32}^{R} = - 0.63~ x_{32}^{R} \sin \theta_L$.

On the other hand, the interaction of the physical Higgs boson with the top quark and the top partner is given by
\begin{equation}
-\mathcal{L} \supset \frac{m_t}{v} \cos^2 \theta_L ~ \bar t_L t_R h + \frac{m_T}{v} \sin^2 \theta_L ~ \bar T_L T_R h +\frac{m_T}{2 v} \sin 2 \theta_L   ~ \bar t_L T_R h+\frac{m_t}{2 v} \sin 2 \theta_L ~ \bar T_L t_R h+\textrm{h.c.}\,.
\end{equation}
We compute the decay rate for $h\to \tau \mu$ with $t$, $T$ and $\Delta_1$ particles running in the loops. After properly including all the diagrams exemplified in Fig.~\ref{fig:LQdiag} (four vertex and four leg diagrams), we find a finite result. For the numerical benchmark, we get
\begin{equation}
y_{\tau\mu}  \approx  \frac{N_c}{16\pi^2} \frac{m_t}{v}~ (0.26 y^{R}_{32}+0.43~ x^{R}_{32} \sin \theta_L)y^{L*}_{3 3} \,,
\end{equation}
where we have linearised in $\sin \theta_L$. Finally, requiring the complete cancelation in $\tau \to \mu \gamma$ (that is, $y_{32}^{R} = - 0.63~ x_{32}^{R} \sin \theta_L$),  we find $y^{R}_{32} y^{L*}_{3 3} = 0.47$ gives the best fit to $h\to \tau \mu$ excess.
In passing we note that if $m_T>m_\Delta$, the $T\to \Delta \ell$ decays can produce spectacular signatures at the LHC.

\section{Conclusions}
\label{sec:conclusions}

Prompted by the recent experimental hint of $h \to \tau\mu$ events by the CMS Collaboration,  we have carefully examined the implications of LFV Higgs decays at the percent level on possible extensions of the SM. In particular we have shown how a tentative $\mathcal B(h\to \tau\mu)$ signal can be combined with other Higgs measurements to yield a robust lower bound on the effective LFV Higgs Yukawa couplings to taus and muons. Then we have reexamined the connection between LFV Higgs decays and LFV radiative decays of charged leptons, and demonstrated using EFT methods that the current CMS hint implies $\tau \to \mu \gamma$ at rates, which could be observable at the Belle II experiment. In explicit models, the $\tau \to \mu \gamma$ constraint is generically much more severe. In fact, an eventual observation of $h \to \tau\mu$ at the LHC together with existing indirect constraints would already single out an extended SM scalar sector as a required ingredient in any natural explanation, the minimal example being THDM of type III. We have also examined purely fermionic SM extensions and models where $h \to \tau\mu$ is generated at loop level, only to show that without the introduction of extra Higgs doublets, reconciling all existing indirect constraints with percent level  $\mathcal B(h\to \tau\mu)$, when at all possible, requires a high degree of fine-tuning. Finally, we have shown how a positive signal of  $h \to \tau\mu$ can be combined with experimental searches for $\mu \to e \gamma$ decays and $\mu - e$ conversions in nuclei to yield robust bounds on $\mathcal B(h \to \tau e)$. In particular, considering only low energy Higgs EFT  effects, the two LFV Higgs decay rates could still be comparable. On the other hand, the THDM III cannot accommodate both respective branching ratios at the percent level. Currently planned improvements in experimental searches for $\mu-e$ LFV processes will be able to probe the product $\mathcal B(h\to \tau \mu) \mathcal B(h\to \tau e)$ at the $10^{-7}$ level in generic EFT and to order $10^{-12}$ or better within the THDM III.

\begin{acknowledgments}
  We thank the authors of Ref.~\cite{Cheung:2015yga} for pointing out an inconsistency in the formulation of the fine-tuning solution with a leptoquark and a vector-like quark in the previous version of this paper.  We acknowledge insightful discussions with Vincenzo Chiochia. This work was supported in part by the Slovenian Research Agency. I.D.\ acknowledges the SNSF support through the SCOPES project No.\ IZ74Z0\_137346 and Croatian Science Foundation support under the project 7118. I.N. acknowledges support in part by the Bundesministerium f\"ur Bildung und Forschung. \end{acknowledgments}

\appendix

\section{Higgs fit}
\label{sec:appHiggsFit}

In the following we explain the method used to fit the LHC Higgs data.
Ideally, we want the measurements to be reported in terms of the signal 
strengths normalized to the SM predictions
\begin{equation}
\mu_{(k)}^{i}=\frac{\sigma_{(k)}}{\sigma_{(k)}^{SM}}\frac{\mathcal{B}_{i}}{\mathcal{B}_{i}^{SM}}\,,
\end{equation}
where index $i$ represents a Higgs decay mode, while $k$ denotes a production channel. 
Such observables could then easily be expressed in terms of new physics parameters. 
However, the actual experimental categories are never pure and contain events from 
different production mechanisms. Furthermore, in order to fully reconstruct the total likelihood 
function, it is necessary to know all the correlations among the different categories, which are available only
to the experimental collaborations.

It has been argued recently in Ref.~\cite{Boudjema:2013qla} that the existing measurements for a 
given decay channel should be presented in terms of two-dimensional likelihoods, in which 
gluon-gluon fusion (ggF) and associated production with a top pair (ttH) are combined as 
one signal ($\mu_{(ggF+ttH)}$), while vector boson fusion (VBF) and associated production 
with a gauge boson (VH) as another, ($\mu_{(VBF+VH)}$). 
This decomposition is particularly useful since, on one hand, ttH is sub-dominant with 
respect to ggF and poorly explored in the present data set, while VBF and VH receive 
common corrections in wide class of NP models obeying the custodial invariance. 
Furthermore, the correlations are automatically provided by the experimental collaborations 
for a given decay channel. 

\begin{table}
\caption{ATLAS Higgs data as used in our fitting procedures. The separation into production mechanisms for a given decay channel is 
		used if provided. \label{tab:Data-used-in-A}}
\begin{centering}
\begin{tabular}{cccc}
\hline\hline
Decay channel & Production mode & Signal strength & Correlation \& Reference\tabularnewline
\hline 
\multicolumn{4}{c}{ATLAS}\tabularnewline
\hline\hline
$h\to b\overline{b}$ & VH & $0.2\pm0.65$ & \cite{TheATLAScollaboration:2013lia}\tabularnewline
\hline
\multirow{2}{*}{$h\to ZZ^{*}$} & ggF+ttH & $1.8\pm0.65$ & \multirow{2}{*}{$\rho=-0.72$,~\cite{Aad:2013wqa}}\tabularnewline
 & VBF+VH & $-0.2\pm3.7$ & \tabularnewline
 \hline
\multirow{2}{*}{$h\to WW^{*}$} & ggF+ttH & $0.82\pm0.37$ & \multirow{2}{*}{$\rho=-0.15$,~\cite{ATLAS-CONF-2014-009}}\tabularnewline
 & VBF+VH & $1.74\pm0.80$ & \tabularnewline
 \hline
\multirow{2}{*}{$h\to\gamma\gamma$} & ggF+ttH & $1.61\pm0.41$ & \multirow{2}{*}{$\rho=-0.28$,~\cite{Aad:2013wqa}}\tabularnewline
 & VBF+VH & $1.87\pm0.80$ & \tabularnewline
 \hline
\multirow{2}{*}{$h\to\tau\tau$} & ggF+ttH & $1.5\pm1.6$ & \multirow{2}{*}{$\rho=-0.55$,~\cite{ATLAS-CONF-2014-009}}\tabularnewline
 & VBF+VH & $1.7\pm0.84$ & \tabularnewline
 \hline
$h\to\textrm{invisible}$ & VH  & $0.13\pm0.31$ & \cite{Aad:2014iia}\tabularnewline
 \hline
 $h\to Z \gamma$ & inclusive  & $2.0\pm4.6$ & \cite{Aad:2014fia}\tabularnewline
 \hline
 $h\to \mu \mu$ & inclusive  & $1.6\pm4.2$ & \cite{ATLAS:2013qma}\tabularnewline
\hline\hline 
\end{tabular}
\par\end{centering}
\end{table}

\begin{table}
\caption{CMS Higgs data as used in our fitting procedures. The separation into production mechanisms for a given decay channel is 
		used if provided. \label{tab:Data-used-in-B}}

\begin{centering}
\begin{tabular}{cccc}
\hline\hline
Decay channel & Production mode & Signal strength & Correlation \& Reference\tabularnewline
\hline 
\multicolumn{4}{c}{CMS}\tabularnewline
\hline\hline 
\multirow{3}{*}{$h\to b\overline{b}$} & VH & $1.0\pm0.5$ & \cite{Chatrchyan:2013zna}\tabularnewline
 & VBF & $0.7\pm1.4$ & \cite{CMS:2013jda}\tabularnewline
 & ttH & $1.0\pm2.0$ & \cite{CMS:2013sea}\tabularnewline
 \hline
\multirow{2}{*}{$h\to WW^{*}$} & ggF+ttH & $0.76\pm0.23$ & \multirow{2}{*}{$\rho=-0.21$, \cite{Chatrchyan:2013iaa}}\tabularnewline
 & VBF+VH & $0.74\pm0.62$ & \tabularnewline
 \hline
\multirow{2}{*}{$h\to ZZ^{*}$} & ggF+ttH & $0.90\pm0.45$ & \multirow{2}{*}{$\rho=-0.69$,~\cite{Chatrchyan:2013mxa}}\tabularnewline
 & VBF+VH & $1.7\pm2.3$ & \tabularnewline
 \hline
\multirow{2}{*}{$h\to\gamma\gamma$} & ggF+ttH & $0.50\pm0.41$ & \multirow{2}{*}{$\rho=-0.50$,~\cite{CMS:ril}}\tabularnewline
 & VBF+VH & $1.64\pm0.88$ & \tabularnewline
 \hline
\multirow{4}{*}{$h\to\tau\tau$} & 0-jet & $0.34\pm1.09$ & \multirow{4}{*}{\cite{Chatrchyan:2014nva}}\tabularnewline
 & 1-jet & $1.07\pm0.46$ &  \tabularnewline
  & 2-jet (VBF-tag) & $0.94\pm0.41$ &  \tabularnewline
  & VH-tag & $-0.33\pm1.02$ &  \tabularnewline
  \hline
  \multirow{3}{*}{$\mathcal{B}(h\to\tau\mu)~[\%]$} & 0-jet & $0.77\pm0.55$ & \multirow{3}{*}{\cite{Khachatryan:2015kon}}\tabularnewline
 & 1-jet & $0.59\pm0.62$ &  \tabularnewline
  & 2-jet & $1.1\pm0.80$ &  \tabularnewline
  \hline
 $h\to\textrm{invisible}$ & VBF+VH  & $0.14\pm0.22$ & \cite{Chatrchyan:2014tja}\tabularnewline
 \hline
 $h\to Z \gamma$ & inclusive  & $0.0\pm4.8$ & \cite{Chatrchyan:2013vaa}\tabularnewline
 \hline
 $h\to \mu \mu$ & inclusive  & $2.9\pm2.8$ & \cite{CMS:2013aga}\tabularnewline
\hline\hline 
\end{tabular}
\par\end{centering}
\end{table}

Following the recommendation, ATLAS and CMS have combined different search categories 
for a given decay mode to provide separation into production mechanisms.
The results are usually presented in 2D plots in ($\mu_{(ggF+ttH)},\mu_{(VBF+VH)}$) plane.
In this case, we parametrize the likelihood with
\begin{equation}
\chi_{1}^{2}=\underset{i}{\sum}\left(\begin{array}{c}
\mu_{(ggF+ttH)}^{i}-\hat{\mu}_{(ggF+ttH)}^{i}\\
\mu_{(VBF+VH)}^{i}-\hat{\mu}_{(VBF+VH)}^{i}
\end{array}\right)^{T}V_{i}^{-1}\left(\begin{array}{c}
\mu_{(ggF+ttH)}^{i}-\hat{\mu}_{(ggF+ttH)}^{i}\\
\mu_{(VBF+VH)}^{i}-\hat{\mu}_{(VBF+VH)}^{i}
\end{array}\right),
\end{equation}
where the sum goes over the decay channels and the covariance matrices are given by
\begin{equation}
V_{i}=\left(\begin{array}{cc}
\left(\hat{\sigma}_{(ggF+ttH)}^{i}\right)^{2} & \rho^{i}\hat{\sigma}_{(ggF+ttH)}^{i}\hat{\sigma}_{(VBF+VH)}^{i}\\
\rho^{i}\hat{\sigma}_{(ggF+ttH)}^{i}\hat{\sigma}_{(VBF+VH)}^{i} & \left(\hat{\sigma}_{(VBF+VH)}^{i}\right)^{2}
\end{array}\right).
\end{equation}
We obtain the best-fit values ($\hat{\mu}$), variances ($\hat{\sigma}$) and correlations
($\rho$) from the plots provided by the experimental collaborations.  These are presented in Tab.~\ref{tab:Data-used-in-A} and Tab.~\ref{tab:Data-used-in-B}. 

If the separation into production modes is not provided, we use the signal strength measurements from existing
search categories. These in general target certain production mechanism, which, however, does not imply $100\%$ purity. 
Inclusive categories are dominated by ggF ($90\%$), while VBF-tagged categories can have $20\%$ to 
$50\%$ contamination from ggF. VH- and ttH-tagged categories are assumed to be pure. In this case, we write
\begin{equation}
\frac{\sigma_{A\to h}}{\sigma_{A\to h}^{SM}}=\xi_{ggF}\frac{\sigma_{ggF}}{\sigma_{ggF}^{SM}}+\xi_{VBF}\frac{\sigma_{VBF}}{\sigma_{VBF}^{SM}}+\xi_{VH}\frac{\sigma_{VH}}{\sigma_{VH}^{SM}}+\xi_{ttH}\frac{\sigma_{ttH}}{\sigma_{ttH}^{SM}},
\end{equation}
where $\xi_{i}$ represent contributions of the specified production mechanisms to the given category. 
We do not consider correlations here and add each measurement to total $\chi^{2}$ separately,
\begin{equation}
\chi_{2}^{2}=\underset{j}{\sum}\left(\frac{\mu_{j}-\hat{\mu}_{j}}{\hat{\sigma}_{j}}\right)^{2}.
\end{equation}
We approximate the likelihood for the total Higgs decay width measurement~\cite{CMS:2014ala} by 
\begin{equation}
\chi^2_{\Gamma}=\left(\frac{\frac{\Gamma_{h}}{\Gamma_{h}^{\rm SM}}-0.3}{1.9}\right)^2.
\end{equation}
In Tab.~\ref{tab:Data-used-in-A} and Tab.~\ref{tab:Data-used-in-B} we finally summarize all the available LHC Higgs data used in our analyses.

The total $\chi^{2}$ function is given by the sum of all the contributions, 
namely, $\chi^{2}=\chi_1^{2}+\chi_2^{2}+\chi^2_\Gamma $. In order to confront the new physics model 
to data, we express all signal strengths ($\mu$) and total Higgs decay width in terms of model parameters and minimize the
$\chi^{2}$ to find the best fit point. The best fit regions are defined by the appropriate cumulative distribution functions, 
namely, $68.2\%$~($1\,\sigma$) best-fit region for one- (two-) parameter fit satisfies $\chi^{2}-\chi_{min}^{2}<1.0~(2.3)$, 
while $95.5\%$~($2\,\sigma$) best-fit region satisfies $1.0~(2.3)<\chi^{2}-\chi_{min}^{2}<4.0~(6.2)$. The other 
parameters in the likelihood are treated as nuisance parameters.

The limits quoted in Eq.~\eqref{eq:htaummu1} are obtained after allowing only $y_{\tau\mu},y_{\mu\tau}$ to be 
free parameter while fixing other Higgs couplings to their SM values. On the other hand, in the limits shown in 
Eq.~\eqref{eq:htaummu1b} we further allow for arbitrary values of the usual SM tree level Higgs
boson couplings; $\kappa_t$, $\kappa_\tau$, $\kappa_b$, $\kappa_V$ and $\kappa_c$ 
where $\kappa_V=\kappa_W=\kappa_Z$
as well as arbitrary new contributions to loop induced Higgs couplings; $\kappa_g$ and $\kappa_\gamma$.

\section{Barr-Zee contributions to $\tau \to \mu \gamma$ in type-III THDM}
The effective coefficients receive contributions due to the Barr-Zee diagrams
\begin{align}
\label{app:BZ}
  c_R^{H\gamma\gamma(t)} &= \frac{3 e^2 G_f Q_t^2}{8 \pi^2}
  \frac{\sqrt{2}m_t}{s_\beta m_\tau} \sum_{k=1,2,3} y^{H_k^0}_{\mu \tau}
  \left[ \mrm{Re} (x^k_u) f(z_{tk}) + i \,\mrm{Im} (x^k_u) g(z_{tk})
  \right]\,,\\
  c_L^{H\gamma\gamma(t)} &= \frac{3 e^2 G_f Q_t^2}{8 \pi^2}
  \frac{\sqrt{2}m_t}{s_\beta m_\tau} \sum_{k=1,2,3} {y^{H_k^0}_{\tau \mu}}^*
  \left[ \mrm{Re} (x^k_u) f(z_{tk}) - i \,\mrm{Im} (x^k_u) g(z_{tk})
  \right]\,,\\
  c_{R}^{H\gamma\gamma(W)} &= \frac{-e^2 G_f}{16 \pi^2} 
  \frac{v}{m_\tau} \sum_{k=H,h} C_{H_k^0 W
    W}  y^{H_k^0}_{\mu \tau} \left[3 f(z_{Wk}) + \frac{23}{4}
    g(z_{Wk})+ \frac{3}{4} h(z_{Wk}) \right]\,,\\
  c_{L}^{H\gamma\gamma(W)} &= \frac{-e^2 G_f}{16 \pi^2} 
  \frac{v}{m_\tau} \sum_{k=H,h} C_{H_k^0 W
    W}  {y^{H_k^0}_{\tau \mu}}^* \left[3 f(z_{Wk}) + \frac{23}{4}
    g(z_{Wk})+ \frac{3}{4} h(z_{Wk}) \right]\,.
\end{align}
Here $C_{H_k^0 WW}$ measure the coupling strength of neutral Higgses to $WW$ relative to the SM coupling $hWW$, i.e., deviation from the vertex $i g m_W g_{\mu\nu}$. Their values are $C_{H WW} = \cos(\beta-\alpha)$, $C_{h WW} = \sin(\beta-\alpha)$.  The coefficients shown above correspond to topologies that contain $H_k\gamma\gamma$ coupling, mediated either by $t$ or $W$ loop. The virtual photon can be replaced by a $Z$ boson, however such contributions are subdominant. Loop integrals $f$, $g$, and $h$ were calculated in Ref.~\cite{Chang:1993kw}, whereas their arguments are defined as $z_{tk} = m_t^2/m_{H_k^0}^2$, and $z_{Wk} = m_W^2/m_{H_k^0}^2$.

\bibliography{refs}
\end{document}